\RequirePackage{ifpdf} 
\ifpdf
\documentclass[pdftex,12pt,a4paper]{article}  
\else
\documentclass[12pt,a4paper]{article} 
\fi

\parskip=\medskipamount

\usepackage[mathscr]{euscript} 	
\usepackage{bbm} 
\usepackage{amsfonts,amsmath,amssymb}
\ifpdf
\usepackage[pdftex]{graphicx}	
\usepackage[usenames,pdftex]{color}		
\usepackage[pdftex,bookmarks]{hyperref}
\hypersetup{
		bookmarksopen,bookmarksnumbered,
		pdfnewwindow=true,
		pdfauthor = {Simon J. Tyler and Sergei M. Kuzenko},
		pdftitle = {Goldstino actions and their symmetries},
		pdfsubject = {},
		pdfkeywords = {Supersymmetry, Goldstino, Akulov-Volkov},
		pdfcreator = {LaTeX with hyperref package},
		pdfproducer = {pdflatex}
}
\else
\usepackage[dvips]{graphicx}
\usepackage[usenames,dvips]{color}
\usepackage[dvips,bookmarks,colorlinks=false]{hyperref}
\fi
\def\a{\alpha}
\def\b{\beta}
\def\c{\chi}
\def\d{\delta}
\def\e{\epsilon}
\def\f{\phi}
\def\g{\gamma}

\def\h{\eta}

\def\j{\psi}
\def\k{\kappa}
\def\l{\lambda}
\def\m{\mu}

\def\q{\theta}

\def\s{\sigma}

\def\x{\xi}

\def\F{\Phi}

\def\L{\Lambda}

\def\X{\Xi}

\newcommand{\vf}{\varphi}
\newcommand{\da}{{\dot{\alpha}}}
\newcommand{\db}{{\dot{\beta}}}

\newcommand{\dsR}{{\mathbb R}}

\def\ds1{\ensuremath{\mathbbm{1}}}
\newcommand{\rmd}{{\rm d}}
\newcommand{\rme}{{\rm e}}
\newcommand{\rmi}{{\rm i}}
\newcommand{\rmr}{{\rm r}}

\newcommand{\cL}{{\cal L}}
\newcommand{\cN}{{\cal N}}
\newcommand{\st}{{\tilde\s}}

\newcommand{\expt}[1]{\ensuremath{\big< #1 \big>}}

\renewcommand{\Re}{\ensuremath{\text{Re}}}
\renewcommand{\Im}{\ensuremath{\text{Im}}}
\newcommand{\pd}{\partial}
 				
\def\tr{{\rm tr}}		

\def\cc {{\rm c.c.}}

\def\intx{\int\!\!{\rmd}^4x\,}

\def\half{\ensuremath{\frac{1}{2}}} 

\newcommand{\be}{\begin{equation}}
\newcommand{\ee}{\end{equation}}
\newcommand{\bea}{\begin{eqnarray}}
\newcommand{\eea}{\end{eqnarray}}
\newcommand{\non}{\nonumber}

\newcommand{\AV}{{\mathrm{AV}}} 
\newcommand{\KS}{{\mathrm{KS}}} 
\newcommand{\SW}{{\mathrm{SW}}} 
\newcommand{\BI}{{\mathrm{BI}}} 
\newcommand{\BG}{{\mathrm{BG}}} 
\newcommand{\oLRa}[1]{\overset{\text{\tiny$\leftrightarrow$}}{#1}}
\newcommand{\Db}{{\bar{D}}}
\newcommand{\Fb}{{\bar{\Phi}}}
\newcommand{\Wb}{{\bar{W}}}
\newcommand{\vb}{{\bar{v}}}
\newcommand{\ub}{{\bar{u}}}
\newcommand{\wb}{{\bar{w}}}
\newcommand{\lt}{{\tilde{\lambda}}}

\newcommand{\vt}{{\tilde v}}
\newcommand{\vbt}{{\bar{\tilde v}}}
\makeatletter
\newcommand*{\Relbarfill@}{\arrowfill@\Relbar\Relbar\Relbar}
\newcommand*{\xeq}[2][]{\ext@arrow 0022\Relbarfill@{#1}{#2}}
\makeatother
\begin{document}                        
\begin{titlepage}

\begin{flushright} February, 2011 \\
Revised version: 18 March, 2011
\end{flushright}

\vspace{5mm}

\begin{center}
{\large \bf  On the Goldstino actions and their symmetries}
\end{center}
\begin{center}
{\large  
{Sergei M.\ Kuzenko}
\footnote{kuzenko@cyllene.uwa.edu.au}
and 
{Simon J.\ Tyler}
\footnote{styler@physics.uwa.edu.au}

\vspace{5mm}

\footnotesize{
{\it School of Physics M013, The University of Western Australia\\
35 Stirling Highway, Crawley W.A. 6009, Australia}}  

\vspace{2mm}}
\end{center}

\vspace{5mm}

\pdfbookmark[1]{Abstract}{abstract_bookmark}
\begin{abstract}
\baselineskip=14pt
Starting from the Akulov-Volkov (AV) action, 
we compute a finite-dimensional Lie group $G$ of all field transformations 
of the form $\l \to {\l}' = \l +O(\l^3)$
which preserve the functional structure of low-energy Goldstino-like actions. 
Associated with $G$ is its twelve-parameter 
subgroup $H$ of trivial symmetries of the AV action. 
The coset space $G/H$ is naturally identified with 
the space of all Goldstino models.
We then apply our construction to study
the properties of five different Goldstino actions available in the literature.
Making use of the most general field redefinition derived,
we find explicit maps between all five cases. In each case 
there is a twelve-parameter freedom in these maps
due to trivial symmetries inherent in the Goldstino actions.
Finally, by using the pushforward of the AV supersymmetry, 
we find the off-shell nonlinear supersymmetry transformations 
of the other five actions and compare to those
normally associated with these actions.
\end{abstract}
\vfill

\end{titlepage}

\tableofcontents{}
\vspace{1cm}
\bigskip\hrule

\section{Introduction}\label{sect:Intro}
The Akulov-Volkov (AV) model \cite{VA} is the second oldest 
supersymmetric theory in four space-time dimensions.
It describes the low-energy dynamics of a massless Nambu-Goldstone spin-1/2
particle which is associated with the spontaneous breaking of rigid 
supersymmetry and  is called the Goldstino. 
A derivation of the AV model using superspace techniques was given in 1973 
\cite{AV} by its discoverers.\footnote%
{The Akulov-Volkov paper \cite{AV}, 
which presented the detailed derivation of the AV model,
was submitted to the journal {\it Theoretical  and Mathematical Physics} 
on 8 January 1973, and published in January 1974. 
The concepts of  the $\cN$-extended super-Poincar\'e group and superspace 
were also introduced in that paper for the first time.  See \cite{K-Srni}
for a recent pedagogical review of the pioneering approach of \cite{AV}.
}
Nice textbook reviews of the AV model are also available, see e.g  \cite{WB}.

Since 1972, a number of variant Goldstino models 
have appeared in the literature.
They can naturally be organised in two different families: 
(i) Goldstino actions formulated in terms
of constrained $\cN=1$ superfields 
\cite{Rocek1978,LR,SamuelWess1983,Casalbuoni1989,KS}; 
(ii) the fermionic sectors of models for partial $\cN=2 \to \cN=1$ 
supersymmetry breaking \cite{BG1,BG2,RT,G-RPR}. 
The models for partial supersymmetry breaking 
which give rise  to (ii) are formulated in terms of constrained 
$\cN=2$ superfields (or unconstrained $\cN=1$ ones). 
Of course, the existence of the zoo of Goldstino actions
does not mean that different models lead to inequivalent dynamics.
Indeed, according to the general theory of the nonlinear realisation of 
${\cal N}=1$ supersymmetry 
\cite{AV,Ivanov1977,Ivanov1978,Ivanov1982,Uematsu:1981rj}, 
the AV action is universal in the sense that 
any Goldstino model should be related to the AV action by a 
nonlinear field redefinition. However, some of 
the variant Goldstino models are interesting in their own right, 
in particular in the context of supergravity \cite{LR,SamuelWess1983},
and therefore it becomes important to work out techniques to construct such 
field redefinitions, which is a nontrivial technical problem. 
So far the explicit construction of required field redefinitions has been 
carried out only for a few Goldstino models
on a case-by-case basis \cite{Rocek1978,KMcC,Liu2010,letter}.
In the present paper, which is an extended version of \cite{letter},
we construct the most general field redefinition relating any Goldstino model 
to the AV action. We also  apply our general method to practically all  of the 
known Goldstino models. 

Our approach is similar in spirit to that of \cite{KMcC} and 
makes no use of the techniques developed within the general theory 
of  spontaneously broken  supersymmetry 
\cite{Ivanov1977,Ivanov1978,Ivanov1982,Uematsu:1981rj},  
and thus it can be applied to more general fermionic theories than the 
Goldstino models.
It particular, it can be employed to study the spin-1/2 sectors of 
supersymmetric Euler-Heisenberg-type actions.
Such nonlinear models originate as low-energy effective actions 
in supersymmetric gauge theories and have the general form:
\begin{align} \label{gendualaction}
  S = \frac{1}{2}  \int {\rm d}^4 x {\rm d}^2\q\, W^2 
	+  \frac{1}{4}\, \int {\rm d}^4 x {\rm d}^4\q\, W^2\,{\bar W}^2  \,
	\L \Big( \frac{1}{8} D^2\,W^2 \, ,\, \frac{1}{8}
	{\bar D}^2\, {\bar W}^2 \Big)\,,
\end{align}
where $W^2 := W^\a W_\a $, 
and $W_\a$ is the chiral field strength of a U(1) vector multiplet. 
The supersymmetric Born-Infeld action \cite{CF,BG1}, 
which will be discussed in section \ref{sect:BI},
is a special representative of this family. 
The corresponding fermionic sector is a variant Goldstino model, 
as was pointed out  in \cite{HK}.
The field redefinition relating the latter action to the AV model 
was constructed in \cite{KMcC}.
In fact, there are infinitely many models of the form \eqref{gendualaction}
such that their fermionic sectors are Goldstino models. 
They belong to the family of models for self-dual nonlinear $\cN=1$
supersymmetric electrodynamics \cite{KT} for which the function  
$\Lambda(u, \bar u)$ in \eqref{gendualaction} obeys 
the following differential equation:
\begin{align} \label{dif}
	\Im  \left\{ \frac{\pd (u \, \L) }{\pd u}
	- \bar{u} \left( \frac{\pd (u \, \L )  }{\pd u} \right)^2 \right\} = 0\;.
\end{align}
As shown in \cite{KMcC},  the fermionic sector
of such theories is equivalent to the  AV model if 
$\L_{u{\bar u}}(0,0) = 3 \L^{3}(0,0)$.%
\footnote{The explanation of this ubiquity of 
Goldstino actions was given in \cite{Kuzenko:2009ym}.}

Some of the results contained in the present paper follow from calculations
that are formidable to do by hand. In particular the results
on the pushforward of the AV supersymmetry over a field redefinition
with all twelve trivial symmetry parameters would be next to impossible%
\footnote{Merely writing out the result of such a calculation would
take as many pages as this entire paper.}
to obtain without computer assistance. 
So, distributed with this paper is the Mathematica \cite{mma} 
notebook created by one of us (SJT)
and used to obtain the results. 
Also included are some data files containing most%
\footnote{%
Some of the data files are very large, so they can not all be included. 
The arXiv only provides limited support for including ancillary files
\url{http://arxiv.org/help/ancillary_files}.
} 
of the results.
The core of the program is the construction of a canonical form for spinor
expressions which then allows for unambiguous comparisons between expressions.
This canonical form is obtained by using an explicit matrix representation of 
the Pauli matrices so that all Fierz identities become trivial upon 
choosing a definite ordering for spinor monomials.
All results are then obtained through the careful  application of
substitution rules.

This paper is organised as follows. 
In section \ref{sect:AV} we use the AV action and the most general
``Goldstino-structure'' preserving field redefinition to generate 
the most general possible Goldstino action. 
This general Goldstino action forms the basis of the rest of our analysis. 
By comparing it with another Goldstino action we can find the map
that takes the AV action to the other Goldstino action.
In this section we also find a twelve parameter group of symmetries 
of the AV action that, in section \ref{sect:Trivialities},  
are shown to be consequences of the trivial symmetries 
present in all Goldstino actions. 
These free parameters show up in all of the maps found in this paper. 
In section \ref{sect:Rocek} we examine the Goldstino action given by Ro\v{c}ek
and generalise his results.
In section \ref{sect:KS} we examine the Komargodski-Seiberg action and
give the map that generates it from the AV action.
This map is then used to generate the model's nonlinear supersymmetry 
transformation, which is hard to obtain using other methods.
Section \ref{sect:SW} 
is devoted to the so-called chiral AV action 
given by Samuel and Wess (and earlier by Zumino). 
This model is related to the normal 
AV action by a simple field redefinition which can be compared
with the more general results found by our method.
Sections \ref{sect:BI} and \ref{sect:CS} study the two type (ii) Goldstino
actions that follow from the three partial supersymmetry breaking Goldstone
actions found in the literature.
Appendices \ref{sect:bases} and \ref{sect:composition}
contain details on the basis used for Goldstino actions in this paper
and the composition and inversion rules for the general field redefinition.
Finally, Appendix \ref{sect:LagMul} is devoted to 
a detailed analysis of the Komargodski-Seiberg Goldstino model 
using the Lagrange multiplier action \eqref{eqn:KS0Lag}.

\section{General Goldstino action}\label{sect:AV}
We start with the AV action \cite{VA}
\begin{align}
\label{eqn:AV1}
	S_\AV[\l,\bar\l] &= \frac1{2\k^2}\intx\Big(1-\det\X\Big)\,,
\end{align}
where $\k$ denotes the 
coupling constant of dimension $({\rm length})^2$ and
\begin{align} \label{defn:Xi v vb}
	\X_a{}^b 
			&= \d_a{}^b+{\k^2}\left(v+\bar{v}\right)_a{}^b\,,&
	v_a{}^b	 &:= \rmi\l\s^b\pd_a\bar\l \,, \quad  \quad 
	\vb_a{}^b := -\rmi\pd_a\l\s^b\bar\l\,.
\end{align}
By construction, $S_\AV $ is invariant under the 
nonlinear supersymmetry transformations
\begin{align} \label{eqn:AV_SUSY}
		\d_\x\l_\a &= \frac1\k\x_\a 
			-\rmi\k \big(\l\s^b\bar\x-\x\s^b\bar\l\big)\pd_b\l_\a \,.
\end{align}
Expanding out the determinant in \eqref{eqn:AV1}
and denoting the trace of a  matrix $M=(M_a{}^b)$ with Lorentz indices as
\( \expt{M} = \tr(M) = M_a{}^a \)
yields
\begin{align} \label{eqn:AV2}
	S_\AV[\l,\bar\l] = -\half&\intx\Bigg(\expt{v+\bar v}
	+2\k^2\Big(\expt{v}\expt{\bar v}-\expt{v\bar v}\Big)\non \\
	&+{\k^4}\Big(\expt{v^2\vb}-\expt{v}\expt{v\vb}
		-\half\expt{v^2}\expt{\vb}+\half\expt{v}^2\expt{\vb}+\cc\Big)\Bigg)\,.
\end{align}
As demonstrated in \cite{KMcC}, 
the 8th-order terms vanish due to an algebraic spinor identity. 

The general structure of $S_\AV$ and any other low-energy Goldstino action is 
schematically%
\footnote{%
Here we ignore higher-derivative corrections to the Goldstino actions.
}
\begin{equation} \label{eqn:GoldStruct}
  S_{\text{Goldstino}} \sim \intx \sum_{n=0}^4  \k^{n-2}\l^n\bar\l^n\pd^n\, .
\end{equation}
This follows from dimensional counting and the fact that
a Goldstino field must parametrize 
a coset space of the $\cN=1$ super-Poincar\'e group
and thus always occur in the combination $\k\,\l= \q$. 
The most general field redefinition that preserves such a structure is
\begin{align} \label{eqn:GeneralFieldRedef}
	\l_\a \to \l'_\a &= \l_\a+{\k^2}\l_\a\expt{\a_1 v + \a_2\vb} 
		+\rmi\a_3{\k^2}(\s^a\bar\l)_\a(\pd_a\l^2)  \\\non  					
	&+{\k^4}\l_\a\big(\b_1\expt{v\vb}+\b_2\expt{v}\expt{\vb}
		+ \b_3\expt{\vb^2} + \b_4 \expt{\vb}^2
		+ \b_5\pd^a\l^2\pd_a\bar\l^2 + \b_6\bar\l^2\Box\l^2\big) ~~~~\\\non
	&+\rmi{\k^4}(\s^a\bar\l)_\a(\pd_a\l^2)\expt{\b_7v +\b_8\vb} \\\non 		
	&+ {\k^6}\l_\a\big(\g_1\expt{v\vb^2}+\g_2\expt{v\vb}\expt{\vb}
		+ \g_3\expt{v}\expt{\vb^2} + \g_4\expt{v}\expt{\vb}^2
		+\g_5\expt{\vb}\pd^a\l^2\pd_a\bar\l^2 \big) \,.~~~~
\end{align}
The coefficients can be complex and we denote their real and imaginary parts as
\begin{align} \label{eqn:CoeffReIm}
	\a_i = \a_i^\rmr + \rmi\a_i^\rmi\,,\quad
	\b_j = \b_j^\rmr + \rmi\b_j^\rmi\,,\quad
	\g_k = \g_k^\rmr + \rmi\g_k^\rmi\ .
\end{align}
This field redefinition is equivalent to that given in \cite{KMcC} up to some 
7-fermion identities.
The proof that it is a minimal basis of all possible terms preserving 
\eqref{eqn:GoldStruct} is provided in the attached Mathematica program.
All Goldstino actions of the form \eqref{eqn:GoldStruct} 
are invariant under rigid chiral  transformations
\begin{equation}
  \l_\a \to \rme^{\rmi \vf  } \l_\a \, ,\qquad 
  \bar\l_{\dot \a} \to \rme^{-\rmi \vf  } \bar\l_{\dot \a}\,.
\end{equation}
Without enforcing this symmetry, 
one can introduce a more general field redefinition than
the one defined by eq.\ \eqref{eqn:GeneralFieldRedef}.

The set of all transformations \eqref{eqn:GeneralFieldRedef} 
forms a 32-dimensional Lie group $G$.
The composition rule for the elements of $G$ is spelled out in Appendix 
\ref{sect:composition}.

By applying the field redefinition \eqref{eqn:GeneralFieldRedef} to 
the AV action we generate the most general Goldstino action.
This can then be compared against other Goldstino actions to find the 
maps that relate them to the AV action.  
The general result, written in the basis of Appendix \ref{sect:bases}, is
\vspace{-10pt}
{\allowdisplaybreaks
\begin{align} \label{eqn:GeneralAV}
	&S_\AV[\l'(\l,\bar\l),{\bar\l'}(\l,\bar\l)] 
	= -\half\intx\Bigg(\expt{v+\bar v}  \non  \\ \non
	&+\k^2\Big[ \big\{(2\a_1+2\a_3+1)\expt{v}^2 - (2\a_3+1)\expt{v^2}+\cc\big\}
		+4\a_ 2^\rmr\expt{v}\expt{\vb}-2\a_3^\rmr\pd^a\l^2\pd_a\bar\l^2\Big] \\
	&+{\k^4} \Big[\Big\{
		(2|\a_3|^2+4\a_2\a_3^*-2\a_2-3\a_3+\a_3^*+2\b_3^*+4\b_6^*-\tfrac12)
			\expt{v^2}\expt\vb  \non \\ \non
	&\qquad-(4|\a_3|^2+8\a_3\a_2^*+4\a_1\a_3^*+4\a_1+4\a_3^\rmr
		  -2\b_1+8\b_6+4\b_8^*+1)\expt{v}\expt{v\vb} \\ \non
	&\qquad+(4|\a_3|^2+4\a_3^\rmr+1)\expt{v^2\vb}
		+(|\a_1|^2+|\a_2|^2+2|\a_3|^2+\a_1\a_2^*+2\a_2^*\a_3\\ \non
	&\quad\qquad+2\a_3^*\a_1+2\a_1+4\a_2^\rmr+6\a_3^\rmr
		+2\b_2+2\b_4^*+2\b_7+2\b_8^* +\tfrac12)	\expt{v}^2\expt\vb  \\ \non
	&\qquad -(2\a_2^*\a_3+\a_1-\a_2^*-2\a_3^*-2\b_5+4\b_6+2\b_8^*)
			\expt{v}\pd^a\l^2\pd_a\bar\l^2  + \cc \Big\}   \\ \non
	&\quad+(|\a_1|^2-|\a_2|^2-4|\a_3|^2
			-8(\a_2^\rmr\a_3^\rmr+\a_2^\rmi\a_3^\rmi+\b_6^\rmr))
			\rmi\l\s^a\bar\l (\expt{v}\oLRa{\pd}_a\expt\vb)\Big] \\ 
	&+{\k^6}\Big[ \Big\{
	  2\big(\a_1(\b_1^*-2\b_5^*-2\a_3)+\a_2^*(3+8\a^\rmr_3-\b_1+2\b_5) 
	  +4\a_3(\a_3^*+\b_5^*) \\\non
	&\quad\qquad +2\a_3^*(3-\b_1+4\b_5+2\b_7+4\a_3^*) 
	  -2\rmi\b_1^\rmi+4\b_5+2\b_7+\g_1\big)\expt{v}\expt{v\vb^2} \\\non
	&\qquad+\big(-\a_2^*(3+2\a_1+\a_2^*-\b_1-2\b_3-4\b_6-2\b_7)
		-2\b_1^*+2\b_3-\a_1\b_1^* \\\non
	&\quad\qquad +2\a_3(1+2\a_1+2\a_3-\b_1^*+2\b_4^*+4\b_6^*+2\b_8^*)
	   	-2\b_4^*+8\rmi\b_6^\rmi \\ \non
	&\quad\qquad -2\a_3^*(2+\a_1+2\a_2^*+2\a_3^*-2\b_3-4\b_6)-2\b_8^*+2\g_3\big)
	  \expt{v}^2\expt{\vb^2} + \cc \Big\} \\\non
	&\quad+2\Re\big(2\a_3(-\b_1^*+2\b_3^*+2\b_5^*+4\b_6^*)
		-\b_1-2\b_3+2\b_5-4\b_6\big)\expt{v^2}\expt{\vb^2} \\\non
	&\quad+4\Re\big(2\a_3(\b_1^*-2\b_5^*-2\a_3-1)+\b_1-2\b_5\big)
		\expt{v\vb v\vb} \\\non
	&\quad-4\Re\big(\a_1(\a_2+2\a_3+2\b_5^*+2\b_7^*)
		+\a_2(3+2\a_2^*-6\a_3-\b_1^*)-\b_1 + \b_2 + 4\b_5\\\non
	&\quad\qquad
		+2\a_3(10\a_2^\rmr+8\a_3^\rmr-\b_1^*+\b_2^*+2\b_5^*+2\b_7^*+3)
		+ 4\b_7 - \g_2 + 2\g_5 \big)
			\expt{v}\expt{\vb}\expt{v\vb} \\\non
	&\quad+2\Re\big(\a_1(2\a_3+\b_2^*+2\b_7^*)
		+\a_2(3+\a_2+4\a_2^*-8\a_3+\b_2^*+2\b_4^*+4\b_6^*+2\b_7^*+2\b_8^*) 
		 \\\non
	&\quad\qquad
		+2\a_3(3+8\a_3^\rmr-\b_1^*+\b_2^*+2\b_4^*+2\b_5^*+4\b_6^*+2\b_7^*+2\b_8^*)
		-\b_1+2\b_2+2\b_4\\\non
	&\quad\qquad+2\b_5+4\b_6+4\b_7+2\b_8+2\g_4) 
		+4(\a_1^\rmr\a_2^\rmr+6\a_2^\rmr\a_3^\rmr+\a_3^\rmr\a_1^\rmr)\big)
		\expt{v}^2\expt{\vb}^2
	\Big]	\Bigg)\,.
\end{align}}%
Note that only the real parts of $\g_4$, $\g_2$ and $\g_5$ occur and that
the latter two only appear in the combination $2\g_5-\g_2$. This corresponds
to the fact that the field redefinitions generated by $\g_2^\rmi$, $\g_4^\rmi$, 
$\g_5^\rmi$ and $\g_5^\rmr=2\g_2^\rmr$ are symmetries of the free action.

The general action \eqref{eqn:GeneralAV} has a nonlinear supersymmetry 
that can be derived from the
pushforward of the AV supersymmetry \eqref{eqn:AV_SUSY}
\begin{align} \label{eqn:AV_SUSY_Pushforward}
\d_\x \l_\a &= \d_\x \l_\a(\l',\bar\l')\Big|_{\l'=\l'(\l,\bar\l)}
		= \d_\x\l'^\b\cdot\frac{\d}{\d\l'^\b}\l_\a(\l',\bar\l')
		+ \d_\x\bar\l'_\db\cdot\frac{\d}{\d\bar\l'_\db}\l_\a(\l',\bar\l')
			\Big|_{\l'=\l'(\l,\bar\l)}\,,
\end{align}
where $\l_\a(\l',\bar\l')$ is the inverse of \eqref{eqn:GeneralFieldRedef}
that can be found using the results of Appendix \ref{sect:composition}.
The explicit, all order expression for this supersymmetry is very long, 
but the leading order is easily calculated
\begin{align} \label{eqn:Gen_SUSY}
	\d_\x \l_\a &= \frac{\x_\a}{\k}
	 + \rmi\k\big((1+2\a_3)(\x\s^a\bar\l)  					
	 	-(1+\a_2)(\l\s^a\bar\x)\big)\pd_a\l_\a  			
	 - \rmi\k\a_1(\x\s^a\pd_a\bar\l)\l_\a	\non \\ 		
	 &\quad - \k \expt{\a_1 v + (\a_2+2\a_3)\vb}\x_\a 		
	 - \frac{\rmi\k}2(\a_2+2\a_3)(\s^a\bar\x)_\a\pd_a\l^2 	
	 + O(\k^3) \ .
\end{align}

By writing the AV action in the basis of Appendix \ref{sect:bases} 
we can compare \eqref{eqn:AV2} with \eqref{eqn:GeneralAV}. 
We find that there is a twelve-dimensional family of
symmetries of the form \eqref{eqn:GeneralFieldRedef}:
\begin{align} \label{eqn:AV2AV}
	\l_\a \to \l'_\a &= \l_\a+\rmi\a_2^\rmi{\k^2}\l_\a\expt{\vb} 
	 +\rmi{\k^4}(\s^a\bar\l)_\a(\pd_a\l^2)\expt{\b_7v +\b_8\vb} \\\non
	&+{\k^4}\l_\a\Big( 2(2\b_6+\b_8^*) \expt{v\vb} 
	 +(4\b_6^\rmr-\b_4^*-\b_7-\b_8^*)\expt{v}\expt{\vb} \\\non
	&- (2\b_6+\rmi\a_2^\rmi)\expt{\vb^2} 
	 + \b_4\expt{\vb}^2
	 + (\tfrac\rmi2\a_2^\rmi+2\b_6+\b_8^*)\pd^a\l^2\pd_a\bar\l^2 
	 + \b_6\bar\l^2\Box\l^2\Big) \\\non
	&+ {\k^6}\l_\a\Big( \g_5\expt{\vb}\pd^a\l^2\pd_a\bar\l^2
	 +(\rmi\a_2^\rmi-2\b_7-4\b_8^\rmr)\expt{v\vb^2} \\\non
	&-(2\a_2^\rmi (2 \b_6^\rmi-\b_8^\rmi)
		+\b_4^\rmr+8\b_6^\rmr-3\b_7^\rmr-\b_8^\rmr
		-\rmi\g_2^\rmi-2\g_5^\rmr)\expt{v\vb}\expt{\vb} \\\non
	&+ (\tfrac\rmi2 \a_2^\rmi(4\b_6+2\b_7+2\b_8^*-1)
		+\b_4^*+2\b_6^\rmr-6\rmi\b_6^\rmi+3\b_8^\rmr+\rmi\b_8^\rmi)
		\expt{v}\expt{\vb^2} \\\non
	&+ (-\tfrac12 \a_2^\rmi(3\b_4^\rmi+4\b_6^\rmi+\b_7^\rmi+3\b_8^\rmi)
		+6\b_6^\rmr-\b_7^\rmr+\rmi\g_4^\rmi)\expt{v}\expt{\vb}^2 \Big) \,,
\end{align}
where, the sake of compactness, we write $\b_6^\rmr=-\tfrac18(\a_2^\rmi)^2$. 
The free parameters in the above field redefinition are
\begin{align} \label{eqn:ReImFreeParams}
	\a_2^\rmi,\; \b_4^\rmr,\; \b_4^\rmi,\; \b_6^\rmi,\; 
	\b_7^\rmr,\; \b_7^\rmi,\; \b_8^\rmr,\; \b_8^\rmi,\; 
	\g_2^\rmi,\; \g_4^\rmi,\; \g_5^\rmr,\; \g_5^\rmi\,.
\end{align}
The set of such transformations is a 12-dimensional subgroup $H$ 
of the group $G$ introduced above. 
In section \ref{sect:Trivialities} we will show that all of the
transformations \eqref{eqn:AV2AV} are \emph{trivial} symmetries.%
\footnote{The definition of a trivial symmetry is given at the beginning of 
section \ref{sect:Trivialities}.}
Such trivial symmetries appear in all of the mappings from one Goldstino
action to another and we will always choose the above set of free parameters.

Although the trivial symmetries \eqref{eqn:AV2AV} 
preserve the structure of the action,
they do not preserve the off-shell form of the nonlinear supersymmetry.
We can restrict the parameters of the 
pushforward supersymmetry \eqref{eqn:AV_SUSY_Pushforward}
to the trivial symmetry parameters of \eqref{eqn:AV2AV}. 
This generates a 12 parameter family of (on-shell equivalent) 
nonlinear supersymmetries for the AV action.
In general, these supersymmetry transformations are quite unwieldy, 
e.g.\ \eqref{eqn:KS_SUSY}, 
but the full result is available in the attached Mathematica program.

\section{\texorpdfstring{Ro\v{c}ek's}{Rocek's} Goldstino action}
\label{sect:Rocek}
According to the general theory of spontaneously broken  supersymmetry 
\cite{AV,Ivanov1977,Ivanov1978,Ivanov1982,Uematsu:1981rj},  
any Goldstino action with nonlinearly realised supersymmetry 
can be derived from a model formulated  in terms of constrained superfields. 
The first paper to provide an explicit superfield construction 
was by Ro\v{c}ek \cite{Rocek1978}. 

It was assumed in \cite{Rocek1978}
 that the Goldstino was 
contained in a chiral superfield with the free action 
\begin{align} \label{eqn:free_WZ_action}
	S[\F, \bar \F]&=\int {\rm d}^4 x {\rm d}^4\q\, 
	\bar\F\F = \intx(\f\Box\bar\f-\rmi\j\pd\bar\j+F\bar F)\,,
\end{align}
where we use the component projections
\begin{align}
 \label{eqn:ComponentProjections-Rocek}
	\F| = \f\,,  \qquad
	D_\a\F| = \sqrt2 \j_\a\,, \qquad
	-\frac14 D^2\F| = F\,.
\end{align}
Ro\v{c}ek then looked for a transformation 
\( \big(\f, \j, F\big) \to \big(\f(\l), \j(\l), F(\l)\big) \)
that mapped the corresponding linear supersymmetry transformation
\begin{align} \label{eqn:FreeSusy}
	\d_\x\f=\sqrt2\x\j\,, \quad
	\d_\x\j_\a=\sqrt2\big(\x_\a F + \rmi(\s^a\bar\x)_\a\pd_a\f\big)\,,\quad
	\d_\x F=-\sqrt2\rmi(\pd_a\j\s^a\bar\x)\,.
\end{align} 
onto the AV supersymmetry transformation \eqref{eqn:AV_SUSY}.
This yielded a unique solution that we reproduce below. 
This solution was then recast in terms of the supersymmetric constraints
\begin{align} 
	\F^2 &=0 \,,   \label{eqn:Roceks_Constraint1} \\
	-\frac14 \F\Db^2\bar\F &= f\F \,, \label{eqn:Roceks_Constraint2}
\end{align}
where $f$ is a dimensional constant inversely proportional to $\k$ 
and is chosen to be real.%
\footnote{The sign of $f$ in the above equation differs from that given 
by Komargodski  and Seiberg \cite{KS}.}

We approach the problem the other way around, i.e.\ we start with the 
free action \eqref{eqn:free_WZ_action} and the constraints 
\eqref{eqn:Roceks_Constraint1} and \eqref{eqn:Roceks_Constraint2}.
We then derive the consequent Goldstino action $S_{\rm R}[\j,\bar\j]$ 
which is compared to the general Goldstino action \eqref{eqn:GeneralAV}
in order to find the map $\l\to\l(\j)$ that takes the AV action to 
$S_{\rm R}$. This map is then inverted to reproduce Ro\v{c}ek's results
$\big(\f(\l), \j(\l), F(\l)\big)$.

As noticed by Ro\v{c}ek (in his discussion of the 2D analogue of the AV model), 
the constraints \eqref{eqn:Roceks_Constraint1} 
and \eqref{eqn:Roceks_Constraint2}
mean that an arbitrary low-energy action
\begin{align}
	S_\mathrm{eff} = \int  {\rm d}^4 x {\rm d}^4\q\, K(\bar\F,\F) +
	\left( \intx {\rm d}^2\q\, P(\F) + \cc \right)\,,
\end{align}
can always be reduced to a functional proportional to the free action. 
Indeed,  provided the first constraint \eqref{eqn:Roceks_Constraint1} holds, 
the action is equivalent, modulo a trivial rescaling of the superfields,  to
\begin{align} \label{effective-action2}
	\tilde{S}_{\rm eff} = \intx\rmd^4\q\, \bar\F\F 
		+ \left( \h \intx\rmd^2\q\,\F + {\rm c.c.}
		\right)\,,
\end{align}
for some constant parameter $\h$. 
Imposing the second constraint \eqref{eqn:Roceks_Constraint2} 
makes all three structures in \eqref{effective-action2} completely equivalent,
so that the action can be written as either a pure kinetic term or 
a pure $F$-term.%
\footnote{In the  approach of Komargodski and Seiberg \cite{KS}, 
which is discussed in the next section,
only the constraint \eqref{eqn:Roceks_Constraint1} is imposed. 
As a result, they work with an action of the form \eqref{effective-action2}.} 

The constraint  \eqref{eqn:Roceks_Constraint1} can be solved explicitly 
in terms of the component fields \cite{Casalbuoni1989,KS}. 
This amounts to the fact that the scalar component of the chiral superfield 
becomes a function of the other fields, 
\begin{align} \label{eqn:SolnTo1stConst}
	\F^2=0 \quad  \iff  \quad \f = \frac{\j^2}{2F} \,.
\end{align}
The second constraint, $\F \bar D^2 \bar \F=-4f\F$,
is used to write the auxiliary field in terms of the spinor.
The simplest approach is to use the highest component of the constraint
to get an implicit equation for $F$
\begin{align} \label{eqn:Fbar1}
	F 
	  =	f + \bar F^{-1}\expt{\ub} 
		- \frac14 \bar F^{-2}\bar\j^2\Box(F^{-1}\j^2)\,.
\end{align}
Here and below we use the notation
\begin{align} \label{eqn:defn of u}
	u=(u_a{}^b)\,, \quad 
	u_a{}^b :=\rmi\j\s^b\pd_a\bar\j \,, \quad \qquad 
	\bar u = (\ub_a{}^b)\,,\quad  
	\ub_a{}^b	:=-\rmi\pd_a\j\s^b\bar\j\,.
\end{align}
We also use the same convention for matrix trace, 
e.g.\ $\expt{\ub}$, as in section \ref{sect:AV}.
Equation \eqref{eqn:Fbar1}
can be solved by repeated substitution. After some work, we find
\begin{align} \label{eqn:F-final}
	F &= f \Big(1 + f^{-2}\expt{\ub} 
		- f^{-4}\big(\expt{u}\expt{\ub} + \frac14\bar\j^2\Box\j^2 \big)
		+ f^{-6}(\expt{u}^2\expt{\ub} + \cc) \non\\
		&\quad + \frac14 f^{-6}\big(\expt\ub\j^2\Box\bar\j^2
			+ 2\expt{u}\bar\j^2\Box\j^2 + \bar\j^2\Box(\j^2\expt\ub)\big) \\
		&\quad -3f^{-8}\big(\expt{u}^2\expt\ub^2
			+\frac14\j^2\bar\j^2\Box(\expt{u}^2-\expt{u}\expt\ub+\expt\ub^2) 
			+ \frac{1}{16}\j^2\bar\j^2\Box\bar\j^2\Box\j^2\big) \Big) \,.\non
\end{align}

To get an action that maps onto the AV action,
we set $S_{\rm R}=-f\intx F$ with $F$ given by \eqref{eqn:F-final} 
and choose $f$ such that $f^{-2}=2\k^{2}$.
This yields
\begin{align} \label{eqn:ActionR}
	S_{\rm R} &= -\frac12\intx\!\Big(\!	\expt{u+\ub}
		+\k^2\big(\pd^a\j^2\pd_a\bar\j^2-4\expt{u}\!\expt\ub\!\big)
		+4\k^4\big(\!\expt{u}\!\big(\bar\j^2\Box\j^2+2\expt{u}\!\expt\ub\!\big)
			+\cc\big) \nonumber\\
		&\quad+24\k^6\big(\expt{u^2}\expt{\ub^2}-3\expt{u}^2\expt\ub^2
			-2\expt{u}\expt\ub\expt{u\ub} 
			-\frac38\j^2\bar\j^2\Box\j^2\Box\bar\j^2\big)
	\Big)\,,
\end{align}
where we have dropped the vacuum energy density
and added surface terms to make $S_{\rm R} $ manifestly real.
This action has a nonlinearly realized supersymmetry that follows from the
linear supersymmetry transformations \eqref{eqn:FreeSusy}
and the solutions to the constraints given above, 
eqs.\ \eqref{eqn:SolnTo1stConst} and  \eqref{eqn:F-final},
\begin{align} \label{eqn:Rocek_NL_Susy}
	\d_\x\j_\a 
	&= \frac1\k\x_\a \Big(1 + 2\k^2\expt{\ub} 
		- 4\k^4\big(\expt{u}\expt{\ub} + \frac14\bar\j^2\Box\j^2 \big)
		+ 8\k^6(\expt{u}^2\expt{\ub} + \cc) \non\\
		&\qquad + 2\k^6\big(\expt\ub\j^2\Box\bar\j^2
			+ 2\expt{u}\bar\j^2\Box\j^2 + \bar\j^2\Box(\j^2\expt\ub)\big) \\
		&\qquad -48\k^8\big(\expt{u}^2\expt\ub^2
			+\frac14\j^2\bar\j^2\Box(\expt{u}^2-\expt{u}\expt\ub+\expt\ub^2) 
			+ \frac{1}{16}\j^2\bar\j^2\Box\bar\j^2\Box\j^2\big)\Big) \non\\
	&\quad + \rmi\k(\s^a\bar\x)_\a\pd_a
		\Big(\j^2
		\Big(1 - 2\k^2\expt{\ub} + 4\k^4\expt{\ub}^2 
			+ \k^4\bar\j^2\Box\j^2\Big)
		\Big)\,. \non
\end{align}

Comparing the action \eqref{eqn:ActionR} to \eqref{eqn:GeneralAV}, 
we find the map that takes $S_\AV$ to $S_{\rm R}$:
\begin{align} \label{eqn:AV2R} \non
	\l_\a &= \j_\a-\k^2(1-\rmi\a_2^\rmi)\j_\a\expt{\ub}
		-\rmi{\k^2}(\s^a\bar\j)_\a(\pd_a\j^2)\Big(\tfrac12					
		-{\k^2}\expt{\b_7 u + \b_8 \ub}	\Big)  \\\non						
	&+{\k^4}\j_\a \Big(2(\rmi\a_2^\rmi+2\b_6+\b_8^*-3)\expt{u\ub}
		+ (1-\tfrac\rmi2\a_2^\rmi-\b_4^*+4\b_6^\rmr-\b_7-\b_8^*)
			\expt{u}\expt{\ub} \\ \non
	&\quad+ (\tfrac32-2\rmi\a_2^\rmi-2\b_6)\expt{\ub^2} 
		+ \b_4 \expt{\ub}^2
		+ (\rmi\a_2^\rmi+2\b_6+\b_8^*-\tfrac52)\pd^a\j^2\pd_a\bar\j^2 
		+ \b_6\bar\j^2\Box\j^2\Big) \\ 
	&+ {\k^6}\j_\a \Big(\expt{u\ub^2}
		- (16 + 2\a_2^\rmi(2\b_6^\rmi-\b_8^\rmi) 
			- 8\b_6^\rmr - 2\b_7^\rmr - 4\b_8^\rmr 
			- \rmi\g_2^\rmi - 2\g_5^\rmr)\expt{u\ub}\expt{\ub} \\ \non
	&\quad- (18 - 2\b_4^* -4\b_6^* - \b_7 + 2\rmi \b_8^\rmi - 4\b_8^\rmr 
			-\rmi\a_2^\rmi(2\b_6+\b_7+\b_8^*-3))\expt{u}\expt{\ub^2} \\ \non
	&\quad- (45 -\b_4^\rmr -12 \b_6^\rmr -\b_8^\rmr -\rmi \g_4^\rmi
			+ \tfrac12\a_2^\rmi(3\b_4^\rmi+4\b_6^\rmi+\b_7^\rmi+3\b_8^\rmi) 
			)\expt{u}\expt{\ub}^2 \\ \non
	&\quad	+ \g_5\expt{\ub}\pd^a\j^2\pd_a\bar\j^2 \Big) \,,  
\end{align}
where $4\b_6^\rmr=5-\tfrac12(\a_2^\rmi)^2$.
The inverse of this map can be found using \eqref{eqn:inversion} and it
only matches the result presented in \cite{Rocek1978} when
the twelve free parameters \eqref{eqn:ReImFreeParams} are set to
\begin{align} \label{eqn:Roceks_coeffs}
	\a_2^\rmi=\b_4=\b_6^\rmi=0\,,\quad
	\b_7=-\frac12\,,\quad 
	\b_8=\frac32\,,\quad
	\g_2^\rmi=\g_4^\rmi=0\,,\quad
	\g_5=-3\,.
\end{align}
By using the composition rules of Appendix 
\ref{sect:composition} it can be checked that all of the extra freedom is due 
to the trivial symmetries of the AV action \eqref{eqn:AV2AV}.

Inverting the above field redefinition with the specific coefficients 
\eqref{eqn:Roceks_coeffs} we obtain the solutions to the constraints 
on the AV side: 
\begin{subequations} \label{eqn:Roceks_Results}
\begin{align} 
	f\f &= \frac12\l^2\left(1+\k^2\expt{\vb}
		+\k^4\bar\l^2(\pd^a\l\s_{ab}\pd^b\l)\right) \,, \\
	f F &= \tfrac1{2}\k^{-2}+\expt{\vb} 
		+\k^2\big(\bar\l^2(\pd^a\l\s_{ab}\pd^b\l)
			-(\l\s_a\bar\l)(\pd^b\l\s_b\pd^a\bar\l)
			-(\pd^b\l\s_a\bar\l)(\l\s_b\pd^a\bar\l)\big)~~~~~~~~ \non \\ 
		&\quad+\tfrac14\k^2\bar\l^2\Box\l^2 
		+\rmi\k^4\bar\l^2\big((\l\s_c\pd^c\bar\l)(\pd^a\l\s_{ab}\pd^b\l)
			-2(\l\s_a\pd^b\bar\l)(\pd^a\l\s_{bc}\pd^c\l)\big) \,,  \\
	\j_\a &= \l_\a + \k^2\l_\a\expt{\vb} + \tfrac\rmi2\k^2(\s^a\bar\l)_\a
		\pd_a\l^2\big(1-\k^2\expt{v-\vb}\big) \non\\ 
		&\quad+\k^4\l_\a\big(\expt{v}\expt\vb-\expt{\vb^2}
			+\tfrac12\pd^a\l^2\pd_a\bar\l^2+\tfrac14\bar\l^2\Box\l^2\big)\\\non
		&\quad+\k^6\l_\a\big(\expt{v\vb^2}-\expt{v\vb}\expt\vb
			-\tfrac12\expt{v}(\expt{\vb^2}-\expt\vb^2)\big)\,.
\end{align}
\end{subequations}
These match Ro\v{c}ek's results (upon setting his parameter  $\a$ to zero) 
up to a couple of small typographical errors in his version of eq.\
(\ref{eqn:Roceks_Results}c).\footnote%
{The calculation with $0\neq\a\in\dsR$ has also been performed 
and the conclusion is identical.
}
Note the absence of any 8-fermion terms in (\ref{eqn:Roceks_Results}b) implies
their absence in the AV action \eqref{eqn:AV2} -- a fact 
rediscovered in \cite{KMcC}.

Now that we have the mapping between $S_\AV$  and $S_{\rm R}$, we can calculate
the pushforward \eqref{eqn:AVtoKS_SUSY_Pushforward} of the AV supersymmetry,
which yields a 12 parameter family of supersymmetry transformations. 
In general, they are quite unwieldy, e.g.\ \eqref{eqn:KS_SUSY}, but
the full result has been calculated and is available in the attached
Mathematica program.
We find that the pushforward of the AV supersymmetry only reduces to the 
supersymmetry \eqref{eqn:Rocek_NL_Susy}
when the free parameters are fixed
to \eqref{eqn:Roceks_coeffs}. 
This explains the uniqueness of Ro\v{c}ek's results.

\section{The Casalbuoni-De Curtis-Dominici-Feruglio-Gatto 
	and Komargodski-Seiberg action}\label{sect:KS}
The action that we analyse in this section 
was introduced by Casalbuoni {\it et al.}\ in 1989 \cite{Casalbuoni1989}. 
Unfortunately, their work has remained largely unnoticed.
The same action has recently been rediscovered 
by Komargodski and Seiberg \cite{KS}. 
The novelty of the Komargodski-Seiberg (KS) approach is, 
in particular, that they related the 
Goldstino dynamics to the superconformal anomaly multiplet 
$X$ corresponding to the Ferrara-Zumino supercurrent \cite{FZ}.
Under the renormalization group, the multiplet of anomalies $X$, 
defined in the UV, flows in the IR to a chiral superfield $X_{NL}$
obeying the constraint $X_{NL}^2=0$.
This type of constraint was first introduced by Ro\v{c}ek \cite{Rocek1978} 
and is discussed in the previous section. 
Finally, one of the crucial results of \cite{KS} is that Komargodski and Seiberg
showed how to generalize their Goldstino action
to include higher-derivative interactions
and couplings to supersymmetric matter.
In this work, as we are only interested in the equivalence of 
the various Goldstino models, we will not consider such interactions. 
The Goldstino model of \cite{Casalbuoni1989,KS} 
will be called the KS action for brevity
(this abbreviation stands for `CDCDFGKS action').
 
The model is described by a single chiral superfield constrained by 
\begin{align}\label{f^2}
	\F^2=0\,. 
\end{align}
As discussed in section \ref{sect:Rocek}, 
the most general low-energy action that can be constructed from $\F$ is
\begin{align} \label{eqn:KS0}
	S[\F,\bar\F] = \intx {\rm d}^4\q\, \bar\F\F 
	+ \left(f\intx {\rm d}^4\q\, \F + \cc\right)\,,
\end{align}
where, without loss of generality, we can choose the coupling constant 
$f$ to be real. As in the previous section, 
we find that for $S_\KS$ to match $S_\AV$ then $f$ must be 
such that $2f^2=\k^{-2}$.
As in Ro\v{c}ek's model, the nilpotent constraint is used to solve for the 
scalar component field \eqref{eqn:SolnTo1stConst}, 
with 
the component fields of $\F$ given in 
\eqref{eqn:ComponentProjections-Rocek}. 
This leaves the component action
\begin{align} \label{eqn:KS0.5}
	S[\j,\bar\j,F,\bar F]
	= \intx \big( -\frac12\expt{u+\ub}
	  + \frac{\bar\j^2}{2\bar F}\Box\frac{\j^2}{2F}
	  + f F + f \bar F + F \bar F
	\big)\,,
\end{align}
where $u_a{}^b$ and ${\bar u}_a{}^b$ are defined in \eqref{eqn:defn of u}.
In Ro\v{c}ek's model the second constraint \eqref{eqn:Roceks_Constraint2} 
is used to eliminate the auxiliary complex field.  
In the KS model one does not have such a constraint, 
and both terms in the action \eqref{eqn:KS0} remain  essential.
The auxiliary complex scalar is removed from \eqref{eqn:KS0.5} 
using its equations of motion, leaving the fermionic action \cite{KS}
\begin{equation}\label{eqn:KS1}
	S_{\KS}[\j,\bar\j] = -\half\intx\Big(\expt{u+\ub}+
		\k^2\pd^a\bar{\j}^2\pd_a \j^2
		+\k^6\j^2\bar\j^2\Box\j^2\Box\bar\j^2\Big)\,,
\end{equation}
Comparing \eqref{eqn:ActionR} with \eqref{eqn:KS1} clearly shows that 
the two actions are different.
The KS action appears to have the simplest form among all the Goldstino models.

In both \cite{Casalbuoni1989} and \cite{KS}, the action \eqref{eqn:KS0} was 
analysed using a Lagrange multiplier field to enforce the constraint $\F^2=0$.
This analysis was used in \cite{KS} to show the \emph{on-shell} equivalence
of the KS action with the Ro\v{c}ek action \eqref{eqn:ActionR}; 
to show the \emph{off-shell} equivalence takes a little more work, 
with the final results presented at the end of section \ref{sect:Trivialities}.
The Lagrange multiplier analysis also has some interesting aspects 
in and of itself 
that warrant the closer examination presented in Appendix \ref{sect:LagMul}.

By writing the KS action in the basis of Appendix \ref{sect:bases} and comparing
with the general action \eqref{eqn:GeneralAV} we find the mapping 
(with $-8\b_6^\rmr = {1 + (\a_2^\rmi)^2}$)
\begin{align} \label{eqn:AV2KS}
	\l_\a &= \j_\a+\rmi\a_2^\rmi{\k^2}\j_\a\expt{\ub} 
		-\tfrac\rmi2{\k^2}(\s^a\bar\j)_\a(\pd_a\j^2)
	+ \rmi{\k^4}(\s^a\bar\j)_\a(\pd_a\j^2)\expt{\b_7u +\b_8\ub} \\\non  					
	&+{\k^4}\j_\a\Big(
	2(\rmi\a_2^\rmi+2\b_6+\b_8^*)\expt{u\ub}
	+ \tfrac12(3-2\b_4^*-2\b_8^*+8 \b_6^\rmr-2\b_7
		-\rmi\a_2^\rmi)\expt{u}\expt{\ub} \\\non
	&- (\tfrac12+2\rmi\a_2^\rmi+2\b_6)\expt{\ub^2} 
	+ \b_4 \expt{\ub}^2
	+ (\tfrac{1}{2}+\rmi\a_2^\rmi+2\b_6+\b_8^*)\pd^a\j^2\pd_a\bar\j^2 
	+ \b_6\bar\j^2\Box\j^2\Big) \\\non
	&+\k^6\j_\a\Big( \expt{u\ub^2}
	+ (2-2\a_2^\rmi(2\b_6^\rmi-\b_8^\rmi)+4\b_6^\rmr+2\b_7^\rmr+2\b_8^\rmr
		+\rmi\g_2^\rmi+2 \g_5^\rmr)\expt{u\ub}\expt{\ub} \\\non
	&+ (1+\rmi\a_2^\rmi(2\b_6+\b_7+\b_8^*-2)+2\b_4^*+2\b_6^\rmr-6\rmi\b_6^\rmi
		+3\b_8^\rmr-\rmi\b_8^\rmi)\expt{u}\expt{\ub^2} \\\non
	&+ (1-\tfrac12\a_2^\rmi(3\b_4^\rmi+4\b_6^\rmi+\b_7^\rmi+3\b_8^\rmi)
		+\tfrac12(\b_4^\rmr+16\b_6^\rmr-\b_7^\rmr+\b_8^\rmr+2\rmi\g_4^\rmi))
		\expt{u}\expt{\ub}^2 \\\non
	&+ \g_5\expt{\ub}\pd^a\j^2\pd_a\bar\j^2 \Big) \,,
\end{align}
that maps the AV action onto the KS action.
By using the composition rules of Appendix \ref{sect:composition} 
it can be checked that all of the freedom in \eqref{eqn:AV2KS}
is due to the trivial symmetries of the AV action.

Alternatively, we can attribute the freedom in \eqref{eqn:AV2KS} 
to the trivial symmetries of $S_\KS$. 
These are found in a similar manner to those of
$S_\AV$ and are given by the field redefinition 
(with  $-8\b_6^\rmr=(\a_2^\rmi)^2$)
\begin{align} \label{eqn:KS2KS}
	\j_\a \to \j'_\a &= \j_\a+\rmi\a_2^\rmi\k^2\j_\a\expt{\ub}  
	+\rmi{\k^4}(\s^a\bar\j)_\a(\pd_a\j^2)\expt{\b_7u +\b_8\ub}\\\non  					
	&+{\k^4}\j_\a\big(
		 2(2\rmi\a_2^\rmi+2\b_6+\b_8^*)\expt{u\ub}
		- (\b_4^*-4\b_6^\rmr+\b_7+\b_8^*)\expt{u}\expt{\ub}\\\non  	
		&- 2(\rmi\a_2^\rmi+\b_6)\expt{\ub^2} 
		+ \b_4 \expt{\ub}^2
		+ (2\rmi\a_2^\rmi+2\b_6+\b_8^*)\pd^a\j^2\pd_a\bar\j^2 
		+ \b_6\bar\j^2\Box\j^2\big) \\\non
	&+ \k^6\j_\a\big( \g_5\expt{\ub}\pd^a\j^2\pd_a\bar\j^2
		+ (2\a_2^\rmi(\b_8^\rmi-2\b_6^\rmi)+\rmi\g_2^\rmi+2\g_5^\rmr)
			\expt{u\ub}\expt{\ub} \\\non
		&+ (2 (\b_4^*+6\b_6^\rmr-2\rmi\b_6^\rmi+\b_8^*)
			+\rmi\a_2^\rmi(2\b_6+\b_7+\b_8^*-4))\expt{u}\expt{\ub^2} \\\non
		&+ (-\tfrac12\a_2^\rmi(3\b_4^\rmi+4\b_6^\rmi+\b_7^\rmi+3\b_8^\rmi)
		+2\b_4^\rmr+12\b_6^\rmr+2\b_8^\rmr+\rmi\g_4^\rmi)\expt{u}\expt{\ub}^2
		 \big) \,.
\end{align}
These symmetries are completely equivalent to those given in \eqref{eqn:AV2AV}.
Due to the simplicity of KS action and its equations of motion, 
it is easiest to prove the triviality of the above transformations rather
than \eqref{eqn:AV2AV}.
This is done in section \ref{sect:Trivialities}. 
To aid this proof, it is convenient to list a generating set of trivial 
transformations obtained by setting all but one of the free parameters to zero. 
The group law follows from the composition rules \eqref{eqn:comp}.
Parametrizing each transformation with some real $\e$,
the 12 symmetries have the coefficients collected in the following table:
\begin{equation} \label{12KS_Syms}
\begin{array}{c|cccccccccccccccc}
   	& \a_1 & \a_2 & \a_3 & \b_1 & \b_2 & \b_3 & \b_4 & \b_5 & \b_6 & 
   		\b_7 & \b_8 & \g_1 & \g_2 & \g_3 & \g_4 & \g_5 \\\hline
  \a_2^\rmr & 0 & \rmi \e  & 0 & \frac{\e(8 \rmi-\e)}{2}   & \frac{-\e^2}{2} & 
  \frac{\e (\e -8 \rmi)}{4}   & 0 & \frac{\e(8 \rmi-\e)}{4}  & 
  \frac{-\e ^2}{8} & 0 & 0 & 0 & 0 & 
  \frac{\e(\e -8 \rmi) (2-\rmi \e )}{4} & \frac{-3 \e^2}{2} & 0 \\\hline
  \b_4^\rmr & 0 & 0 & 0 & 0 & -\e  & 0 & \e  & 0 & 0 & 0 & 0 & 0 &
  	 0 & 2 \e  & 2 \e  & 0 \\
  \b_4^\rmi & 0 & 0 & 0 & 0 & \rmi \e  & 0 & \rmi \e  & 0 & 0 & 0 &
  	 0 & 0 & 0 & -2 \rmi \e  & 0 & 0 \\
  \b_6^\rmi & 0 & 0 & 0 & 4 \rmi \e  & 0 & -2 \rmi \e  & 0 &
  	 2 \rmi \e  & \rmi \e  & 0 & 0 & 0 & 0 & -4 \rmi \e  & 0 & 0 \\
  \b_7^\rmr & 0 & 0 & 0 & 0 & -\e  & 0 & 0 & 0 & 0 & \e  & 0 & 0 & 0 &
  	 0 & 0 & 0 \\
  \b_7^\rmi & 0 & 0 & 0 & 0 & -\rmi \e  & 0 & 0 & 0 & 0 & \rmi \e  &
  	 0 & 0 & 0 & 0 & 0 & 0 \\
  \b_8^\rmr & 0 & 0 & 0 & 2 \e  & -\e  & 0 & 0 & \e  & 0 & 0 & \e  & 0 &
  	 0 & 2 \e  & 2 \e  & 0 \\
  \b_8^\rmi & 0 & 0 & 0 & -2 \rmi \e  & \rmi \e  & 0 & 0 & -\rmi \e  &
  	 0 & 0 & \rmi \e  & 0 & 0 & -2 \rmi \e  & 0 & 0 \\\hline
  \g_2^\rmi & 0 & 0 & 0 & 0 & 0 & 0 & 0 & 0 & 0 & 0 & 0 & 0 & \rmi \e  &
  	 0 & 0 & 0 \\
  \g_4^\rmi & 0 & 0 & 0 & 0 & 0 & 0 & 0 & 0 & 0 & 0 & 0 & 0 & 0 & 0 &
  	 \rmi \e  & 0 \\
  \g_5^\rmr & 0 & 0 & 0 & 0 & 0 & 0 & 0 & 0 & 0 & 0 & 0 & 0 & 2 \e  & 0 &
  	 0 & \e  \\
  \g_5^\rmi & 0 & 0 & 0 & 0 & 0 & 0 & 0 & 0 & 0 & 0 & 0 & 0 & 0 &
  	 0 & 0 & \rmi \e 
 \end{array}
\end{equation}
The last four symmetries (associated with the free $\g_i$ parameters) are
trivial symmetries of the free Majorana spinor action.

${}$Finally, we come to the question of  how the supersymmetry algebra is
realised on the fields $\j_\a$ of the KS action.  
The original off-shell linear
supersymmetry of the chiral field $\F$ becomes nonlinear 
in the action \eqref{eqn:KS0.5} and is only realised on-shell after 
the auxiliary equations of motion are enforced to yield \eqref{eqn:KS1}.
This was discussed in \cite{Casalbuoni1989} and the structure of 
an \emph{off-shell} nonlinearly realised supersymmetry that $S_\KS$ 
should possess has been an open question since. 
Now, more than twenty year later, we are in a position to address the problem!

We can calculate the pushforward of the AV supersymmetry
to give us the supersymmetry enjoyed by the KS action
\begin{align} \label{eqn:AVtoKS_SUSY_Pushforward}
\d_\x \j_\a &= \d_\x \j_\a(\l,\bar\l)\Big|_{\l=\l(\j,\bar\j)}
		= \d_\x\l^\b\cdot\frac{\d}{\d\l^\b}\j_\a(\l,\bar\l)
		+ \d_\x\bar\l_\db\cdot\frac{\d}{\d\bar\l_\db}\j_\a(\l,\bar\l)
			\Big|_{\l=\l(\j,\bar\j)}\,,
\end{align}
where $\j=\j(\l,\bar\l)$ and $\l=\l(\j,\bar\j)$ are exact inverses.
In \cite{letter} we used the map obtained from \eqref{eqn:AV2KS} by setting
all twelve free parameters to zero and its inverse calculated using 
\eqref{eqn:inversion}, to find the leading order terms to the KS supersymmetry.
By generating a basis for all possible supersymmetry terms, 
which is available and proved to be minimal in the attached Mathematica program,
it was possible to automate the rest of the calculation. 
The full result is
{\allowdisplaybreaks
\begin{align}\label{eqn:KS_SUSY} 
	\d_\x\j_\a &= \frac1\k \x_\a \non
	+ \k\Big( 
		\x_a\expt{\ub} - (\rmi\j\s^a\bar\x)\pd_a\j_\a
		+\frac12(\rmi\s^a\bar\x)_\a\pd_a\j^2 \Big) 	
		\\\non
	&+ \k^3\Big( 
		\x_\a\big\{
			  \frac12\expt{u\ub} - \expt{u}\expt{\ub}
			- \frac14\expt{\ub^2} 
			- \frac14 \pd^\a\j^2\pd_a\bar\j^2 
			+ \frac18 \bar\j^2\Box\j^2 \big\}\\\non
	&\quad	+ \j_\a\big\{\Box\j^2\bar\x\bar\j 
			- \frac32\pd^a\j^2\pd_a\bar\x\bar\j 
			- \frac32(\j\Box\j)\bar\x\bar\j + \frac34\Box(\j\x)\bar\j^2
			- \frac12\pd^a(\j\x)\pd_a\bar\j^2 \\ 
	&\qquad	+ \frac12(\x\s^a\pd_b\bar\j)(\pd_a\j\s^b\bar\j) 
			+ (\x\s^a\bar\j)(\pd_b\j\s^b\pd_a\bar\j)\big\} \\\non
	&\quad	+ \pd_a\j_\a\big\{
			  (\rmi\j\s^a\bar\x)\expt{\tfrac12\ub-u}
			+ (\j\s^a\pd_b\bar\j)(\x\s^b\bar\j)
			+ (\x\s^a\pd_b\bar\j)(\j\s^b\bar\j)\\\non
	&\qquad	- \frac34 (\rmi\s^a\bar\x)_\a\pd_a\j^2\expt{\ub} 
			- \frac34(\s^a\pd_b\bar\j)_\a\pd_a\j^2(\j\s^b\bar\x)
			- \j_\a(\pd_a\pd_b\j\s^a\bar\x)(\j\s^b\bar\j)\big\}
			\Big)\allowdisplaybreaks\\\non
	&+\k^5\Big( 
		\x_a\big\{\pd^a\j^2\pd_a\bar\j^2\expt{\tfrac34\ub-u}
			+ \frac18\expt{\ub}\j^2\Box\bar\j^2 
			- \frac18\expt{u}\bar\j^2\Box\j^2  
			- \expt{u}^2\expt{\ub}	\\\non
	&\qquad	+ \expt{u\ub}\expt{\tfrac32\ub-u}
			- \frac34\expt{v^2}\expt{\vb} - \frac14\expt{v}\expt{\vb^2}
			+ \frac12\expt{v^2\vb} + \frac12\expt{v\vb^2}
			\big\}\\\non
	&\quad+ \j_\a\big\{
			  \frac14\expt{u}\bar\j\bar\x\Box\j^2
			+ 4\expt{\ub}\pd^b\j^2\pd_b\bar\j\bar\x
			+ \frac34(\rmi\j\s^a\bar\x)\Box\j^2\pd_a\bar\j^2
			 \\\non
	&\qquad	- \frac12 (\rmi\pd_a\j\s^a\bar\x)\big(
			  \expt{u}\expt{\ub} + \expt{u\ub}
			+ \frac34\bar\j^2\Box\j^2 - \frac12(\pd\j)^2\bar\j^2 
			+ \frac32\pd^b\j^2\pd_b\bar\j^2\big) \\\non
	&\qquad	- \frac12 (\rmi\x\s^a\pd_a\bar\j)\big( 
			  \expt{u\ub}
			+ \frac32(\pd\j)^2\bar\j^2 
			+ \pd^b\j^2\pd_b\bar\j^2 \big)\\ \non 
	&\qquad	+ \frac14(\rmi\x\s^a\bar\j)\big( 
			  \pd_a\j^2\Box\bar\j^2
			- \pd_a\j^2(\pd\j)^2	
			- \pd_a\pd_b\j^2\pd^b\bar\j^2	\big) \\\non
	&\qquad+	(\rmi\x\s^a\pd_b\bar\j)\big( 
			- \frac14\pd^b\j^2\pd_a\bar\j^2
			- (\pd_a\j\s^c\pd_c\bar\j)(\j\s^b\bar\j)
			+ \frac12(\pd_a\pd_c\j\s^b\bar\j)(\j\s^c\bar\j) \big)\\\non
	&\qquad+ \expt{\ub}\big(
			  3(\pd_b\j\s^a\bar\x)(\j\s^b\pd_a\bar\j)
			+ 2(\pd_a\pd_b\j\s^a\bar\x)(\j\s^b\bar\j) \big)
			\big\}\\\non
	&\quad+\pd_a\j_\a\big\{ \frac12(\rmi\j\s^a\bar\x)\big(
			  \expt{u}\expt{\ub} + \expt{u\ub}
			- 3\expt{\ub}^2 - \frac54\bar\j^2\Box\j^2 
			- (\pd\j)^2\bar\j^2 \big) \\\non
	&\qquad	+ \frac12(\rmi\x\s^a\bar\j)\expt{u\ub} 
			+ \frac12(\rmi\pd_b\j\s^a\bar\x)(\j\s^b\pd_c\bar\j)(\j\s^c\bar\j)  
			- \frac74(\rmi\j\s_b\bar\x)\bar\j^2\pd^a\pd^b\j^2\\\non
	&\qquad	+ \frac12(\rmi\x\s^c\pd_b\bar\j)\big(
			  (\pd_c\j\s^b\bar\j) - (\j\s^b\pd_c\bar\j)\big)(\j\s^a\bar\j)
			\big\}
			+ \frac32\pd_a\pd_b\j_\a (\rmi\pd^b\j\s^a\bar\x)\j^2\bar\j^2 \\\non
	&\quad	- \frac14(\rmi\s^a\bar\j)_\a\j^2\pd^b(\j\x)\pd_a\pd_b\bar\j^2 
			-\frac18(\rmi\s^c\st^b\pd_b\j)_\a\j^2\big(
			  \bar\j^2(\x\s^a\pd_a\pd_c\bar\j)
			+ 2(\pd_a\bar\j\pd_c\bar\j)(\x\s^a\bar\j)\big)
	\Big)\allowdisplaybreaks\\\non
	&+ \frac18\k^7\Big( 
	\x_\a\big\{2\expt{u^2\ub^2} + 13\expt{(u\ub)^2}
			- 10\expt{u}\expt{u\ub^2} - 4\expt{u^2\ub}\expt\ub\\ \non 
	&\qquad	- 5\expt{u^2}\expt{\ub^2} + 5\expt{u\ub}^2
	 		+ \frac{11}4\j^2\bar\j^2\pd^a\pd^b\j^2\pd_a\pd_b\bar\j^2 \big\} \\\non
	&\quad+ \j^2\bar\j^2\big\{ 
			  \frac14(\s^b\st^a\pd_b\j)_\a\pd_a(\j\x)\Box\bar\j^2
			+ \pd^b\j_\a(\x\s^c\pd_c\bar\j)\big(
			  (\pd_a\j\s^a\pd_b\bar\j)
			+ 6(\pd_b\j\s^a\pd_a\bar\j) \big) \\\non
	&\qquad	+ 5(\s^a\pd_b\bar\j)_\a\big(
			  (\pd_a\j\s^b\pd_c\bar\j)\pd^c(\j\x)
			- (\pd_a\j\s^c\pd_c\bar\j)\pd^b(\j\x) \big)\\\non
	&\qquad	+ \frac54(\s^a\st^c\x)_\a\Box(\pd_a\j^2\pd_c\bar\j^2)
			+ \frac72(\s^c\pd_c\bar\j)_\a\pd_a\pd_b\j^2(\x\s^a\pd^b\bar\j)\\\non
	&\qquad	- \pd_b\j_\a\big(
			  5(\pd_a\j\s^d\st^c\s^a\pd_d\bar\j)(\x\s^b\pd_c\bar\j)
		  	- 4(\pd_a\j\s^d\st^b\s^a\pd_d\bar\j)(\x\s^c\pd_c\bar\j)
		  	\big)\\\non
	&\qquad  -\frac12\pd_a\j_\a\big(
			  13\pd^a(\bar\j\bar\x)\Box\j^2
			- 37\Box(\j\s^a\bar\j)(\pd_b\j\s^b\bar\x) \big)
			- 16(\s^b\bar\x)_\a\pd_a\pd_b\j^2  (\pd_c\j\s^c\pd^b\bar\j) \\\non
	&\qquad	+\frac{37}{2}\pd_b\pd_c\j^2\big(
			  (\s^b\st^a\pd_a\j)_\a\pd^c(\bar\j\bar\x)
			- (\s^b\st^a\pd^c\j)_\a\pd_a(\bar\j\bar\x) \big) \\\non
	&\qquad	+ \pd_a\pd_b\j^2\big(
			  4(\s^c\bar\x)_\a(\pd^b\j\s^a\pd_c\bar\j)
			+ 24(\s^a\bar\x)_\a(\pd^b\j\s^c\pd_c\bar\j) \big)\\\non
	&\qquad  - 15(\s^b\st^c\pd_a\j)_\a (\pd_c\j\s^a\pd_d\bar\j)(\pd_b\s^d\bar\x)
	\big\}\Big)\,.
\end{align}}%
So we see that the cost of the simple action $S_\KS$ is the 
complicated supersymmetry transformation $\d_\h\j_\a$. 
When using the above basis 
(which is probably suboptimal for describing the KS supersymmetry), 
there does not seem to be much simplicity to be gained 
in choosing different trivial symmetry parameters in \eqref{eqn:AV2KS}.
The full 12-parameter family of KS supersymmetry transformations 
is available in the attached Mathematica program, 
but the structure is too unwieldy to reproduce here.
It has explicitly been checked that this mapping satisfies 
the supersymmetry algebra and leaves the action invariant.

The rest of the Goldstino actions considered in this paper have a natural 
nonlinear supersymmetry that is either the starting point for the model
or follows from the combination of a linear supersymmetry 
and some supersymmetric constraints. 
This means that for these actions there is a specific choice of the 
12 trivial symmetry parameters that allows for an organising and simplifying 
of the nonlinear supersymmetry. It is not clear if such set of parameters
and consequent simplification can be found for the supersymmetry of the 
KS action.

In this paper, we only examine the minimal KS action. 
In \cite{KS}, Komargodski and Seiberg showed 
how to use constrained superfield methods to easily construct 
Goldstino models with higher derivative and matter couplings.
This is where the advantage of their approach becomes apparent. 
The only drawback of their approach is that 
after the elimination of the auxiliary field contained in $\F$, 
the original  supersymmetry only closes on-shell.
This does not mean that their models do not possess 
an off-shell nonlinearly realized  supersymmetry.
Provided there exists an invertible field redefinition that takes 
their non-minimal actions to one of the standard nonlinear realizations 
that support higher derivative and matter couplings,
e.g.\ \cite{AV,SamuelWess1983,Ivanov1977,Ivanov1978,Ivanov1982,ClarkLove}, 
then an off-shell supersymmetry may be found using the
push-forward method presented above.

\section{The chiral Alkulov-Volkov action}\label{sect:SW}
The AV supersymmetry transformation  \eqref{eqn:AV_SUSY} 
mixes the fields $\l$ and $\bar\l$.  It was
Zumino \cite{Zumino:ChiralNLSusy} who
introduced an alternate form of nonlinearly realised  
supersymmetry that does not have such a mixing
\begin{align} \label{eqn:chAV_SUSY}
	\d_\x\tilde\l_\a = 
		\frac1\k\x_\a - 2\rmi\k(\tilde\l\s^a\bar\x)\pd_a\tilde\l_\a \,.
\end{align}
This lack of mixing simplifies many types of calculations, 
a fact that was first noticed and exploited
by Samuel and Wess \cite{SamuelWess1983}.
This new supersymmetry is related to the AV one via the simple 
field redefinition \cite{Ivanov1978,SamuelWess1983}
\begin{align} \label{eqn:AVtoChAV_Simp}
	\tilde\l_\a(x) 
	= \l_\a(y)\,, \qquad 
		y=x-\rmi\k^2\l(y)\s\bar\l(y)\,,
\end{align}
which is essentially a nonlinear version of the relations defining 
the chiral superspace coordinates.
The above field redefinition has been explicitly expanded many times in 
the literature and the result can be written in terms of the general
field transformation \eqref{eqn:GeneralFieldRedef} with the parameters
\begin{equation}\label{eqn:AVtoChAV_Simp_Params}\begin{gathered} 
	\a_1=\b_1=\b_3=\g_5=0\,, \quad
	\a_2=\b_2=-\b_4=-\g_1=\g_2=-1\,, \\
	\a_3=-\b_5=\b_7=-\b_8=-\g_3=\g_4=1/2\,, \qquad
	\b_5=-1/4\,.
\end{gathered}\end{equation}

The action for this model is normally constructed in terms of  the superfield
\begin{gather} \label{eqn:chAV_Superfield}
	\tilde\L_\a(x,\q,\bar\q) = \exp(\q Q+\bar\q\bar Q)\tilde\l_\a(x)\,,
\end{gather}
where $(\q Q+\bar\q\bar Q)\lt_\a=\d_\q\lt_\a$ is the transformation 
\eqref{eqn:chAV_SUSY} using the parameter $\q$ instead of $\x$.
The action is then
\begin{equation}\label{eqn:chAV_Action}\begin{aligned} 
	S_\SW &= -\frac{\k^2}2\intx \rmd^4 \q\tilde\L^2\bar{\tilde\L}^2
		= -\frac{\k^2}2\intx\rmd^4\q\frac1{4!}\d_\q^4
			(\tilde\l^2\bar{\tilde\l}^2) \\
		&=-\frac12\intx\Big(\k^{-2}+\expt{\vt+\vbt}
			+\k^2\big(\pd^a\lt^2\pd_a\bar\lt^2+4\expt{\vt}\expt{\vbt}\big) \\
		&\qquad	+\k^4\big(\expt{\vt}\big(2\pd^a\lt^2\pd_a\bar\lt^2
				+4\expt{\vt\vbt}+4\expt{\vbt}^2-2\expt{\vbt^2}
				-\bar\lt^2\Box\lt^2\big) + \cc\big)	\\
		&\qquad	+\k^6\big(\lt^2\bar\lt^2\Box\lt^2\Box\bar\lt^2
				-8\expt{\vt}^2\expt{\vbt^2}-8\expt{\vt^2}\expt{\vbt}^2\big)
		\Big)\,.
\end{aligned}\end{equation}
A similar superfield approach can also be used to reproduce the normal 
AV action \eqref{eqn:AV2} \cite{WB}.
In \cite{mCL} we show how this action and superfield have 
an equivalent description in terms of a constrained complex linear superfield 
that is, in some ways, the more fundamental object.

We can also use the superfield \eqref{eqn:chAV_Superfield} 
to solve Ro\v{c}ek's constraints 
\eqref{eqn:Roceks_Constraint1} and \eqref{eqn:Roceks_Constraint2}.
Following \cite{SamuelWess1983}, it can be shown that  
\begin{align} \label{eqn:SW_Soln_of_Roceks_Constraints}
	\F  = -\frac{\k^2}{8f} \Db^2(\tilde\L^2\bar{\tilde\L}^2)
\end{align}
solves both constraints 
\eqref{eqn:Roceks_Constraint1} and \eqref{eqn:Roceks_Constraint2} 
and immediately gives the relationship between
Ro\v{c}ek's model and the chiral AV Goldstino.
A similar construction starting with the normal AV Goldstino can be used
to reproduce \eqref{eqn:Roceks_Results} with minimal effort.
This approach is related to the general approach \cite{Ivanov1978} 
based on nonlinear representation theory 
and will not be further investigated in this paper.

By writing the action \eqref{eqn:chAV_Action} in the basis of Appendix 
\ref{sect:bases} and comparing to \eqref{eqn:GeneralAV}, 
we find the map that takes $S_\AV$ to $S_\SW$ 
(with $-8\b_6^\rmr = 2+(\a_2^\rmi)^2$):
\begin{align} \label{eqn:AV2SW}
	\l_\a &= \lt_\a+{\k^2}(1+\rmi\a_2^\rmi)\lt_\a\expt{\vbt} 
		-\frac\rmi2 {\k^2}(\s^a\bar\lt)_\a(\pd_a\lt^2)\Big( 1						
		-2{\k^2}\expt{\b_7 \vt + \b_8 \vbt}	\Big)  \\\non						
	&+{\k^4}\lt_\a\big(2(1+\rmi\a_2^\rmi+2\b_6+\b_8^*)\expt{\vt\vbt}
		+ (2-\tfrac\rmi2\s_2^\rmi-\b_4^*+4\b_6^\rmr-\b_7-\b_8^*)
			\expt{\vt}\expt{\vbt} \\ \non
		&\qquad- (\tfrac12+2\rmi\a_2^\rmi+2\b_6)\expt{\vbt^2} 
		+ \b_4 \expt{\vbt}^2
		+ (\tfrac32+\rmi\a_2^\rmi+2\b_6+\b_8^*)\pd^a\lt^2\pd_a\bar\lt^2 
		+ \b_6\bar\lt^2\Box\lt^2\big) \\\non
	&+ {\k^6}\lt_\a\big(\expt{\vt\vbt^2}
		+ (2\b_7^\rmr+\rmi\g_2^\rmi+2\g_5^\rmr-2\a_2^\rmi(2\b_6^\rmi-\b_8^\rmi))
			\expt{\vt\vbt}\expt{\vbt} \\\non
		&\qquad- (2-2\b_4^*+8\rmi\b_6^\rmi+\b_7-2\b_8^\rmr 
			+ \rmi\a_2^\rmi(1-2\b_6-\b_7-\b_8^*))\expt{\vt}\expt{\vbt^2} \\\non
		&\qquad+ (1+4\b_6^\rmr-\b_7^\rmr - \tfrac12\a_2^\rmi(3\b_4^\rmi
			+4\b_6^\rmi+\b_7^\rmi+3\b_8^\rmi)+\rmi\g_4^\rmi)
			\expt{\vt}\expt{\vbt}^2
		+ \g_5\expt{\vbt}\pd^a\lt^2\pd_a\bar\lt^2 \big) \,.
\end{align}

Using this map we can then calculate the pushforward of the AV supersymmetry.
As expected, it only matches \eqref{eqn:chAV_SUSY} for a single choice of the 
free parameters:
$\b_4^\rmr=1$, $\b_7^\rmr=\b_8^\rmr=-1/2$ 
with all other free parameters are zero. 
This choice of parameters also 
reduces \eqref{eqn:AV2SW} back to the inverse of 
\eqref{eqn:AVtoChAV_Simp}.

\section{The supersymmetric Born-Infeld action}\label{sect:BI}
The  $\cN=1$ supersymmetric Born-Infeld (SBI) action
was  introduced for the first time in Refs.\  \cite{CF,DP}
as a supersymmetric extension of the Born-Infeld theory \cite{BornI},
and as such it is not unique.
Bagger and Galperin \cite{BG1}, and  later Ro\v{c}ek and Tseytlin \cite{RT}, 
using alternative techniques, discovered that the action 
given in \cite{CF} describes a Goldstone-Maxwell multiplet
associated with partial  $\cN=2 \to \cN=1$ supersymmetry breaking.
Although this Goldstone action was argued to be unique \cite{BG1,RT}, 
there exists, in fact, 
a two-parameter deformation of the theory \cite{Kuzenko:2009ym} which also 
describes partial  $\cN=2 \to \cN=1$  supersymmetry breaking.
The SBI action is known to be invariant under U(1) duality rotations 
\cite{KT,BMZ}.

The most elegant way to formulate the SBI theory is as the vector Goldstone 
action for partially broken $\cN=2$ supersymmetry \cite{BG1,RT}.
The approaches developed in \cite{BG1} and \cite{RT} are rather 
different from the conceptual point of view.
Either way, the result is a manifestly $\cN=1$ supersymmetric nonlinear theory 
of an Abelian vector multiplet. Its action is given 
in terms of a constrained chiral superfield $X$ constructed in terms of the 
vector-multiplet field strength 
$W_\a$ and its conjugate ${\bar W}_{\dot \a}$
\begin{align} \label{eqn:BI-constrained}
	S[W,\bar W] = \frac14\int  {\rm d}^4 x {\rm d}^2\q\, X + \cc\,, \qquad
	X + \frac{\k^2}4 X \bar D^2 \bar X = W^2\,.
\end{align}
The constraint is solved by \cite{BG1}
\begin{align} \label{eqn:BI-constraint-soln}
\begin{gathered}
	X = W^2 - \frac{\k^2}{2}\bar D^2 (W^2\bar W^2 f(A,B))\,,\quad
	f(A,B)^{-1} = 1+\frac12A+\sqrt{1+A+\frac14B^2}\,,\\
	A = \frac{\k^2}2(D^2W^2+\Db^2\Wb^2)\,,\quad
	B = \frac{\k^2}2(D^2W^2-\Db^2\Wb^2)\,.
\end{gathered}\end{align}
This gives the SBI action
\begin{align} \label{eqn:BI-Susy}
	S[W,\bar W] = 
	\frac{1}{2}  \int  {\rm d}^4 x {\rm d}^2\q\, W^2 
		+ \k^2\int  {\rm d}^4 x {\rm d}^4\q\, W^2\Wb^2 f(A,B)\,. 
\end{align}
The action is also invariant under the nonlinearly realised 
(non-manifest) $\cN=2$ supersymmetry transformation
\begin{align} \label{eqn:BI-SUSY0}
	\d_\h W_\a = \frac1\k \left(\h_\a + \frac{\k^2}4\h_\a\Db^2\bar X 
		+ \rmi\k^2(\s^a\bar\h)_\a\pd_aX \right)\,, \quad
	\d_\h X = \frac2\k \h^\a W_\a\,.
\end{align}

Projection to the fermionic action is consistent with both the equations
of motion and the second supersymmetry \cite{KMcC}. We use the component projections 
\begin{align} \label{eqn:BI-project}
	W_\a| = \c_\a\,,\quad 
	\frac1{2\rmi}D_{(\a}W_{\b)}|=F_{\a\b}\to0 \,, \quad
	-\frac12 D^\a W_\a| = D \to 0\,,
\end{align}
to find the Goldstino action
\begin{align} \label{eqn:BI}
	S_\BI[\c,\bar\c] = -\frac12&\intx\Big(	\expt{w+\bar w}
	+ \k^2\big(\pd^a\c^2\pd_a\bar\c^2 - 4\expt{w}\expt{\wb}\big) \\\non
	&+ 8\k^4\big(\expt{w}^2\expt\wb+\frac12\expt{w}\bar\c^2\Box\c^2+\cc\big) 
		-12\k^6\big(\expt{w}^2\bar\c^2\Box\c^2 + \cc \big) \\\non
	&- 48\k^6\big(\expt{w}^2\expt{\wb}^2 
		-\frac12\expt{w}\expt{\wb}\pd^a\c^2\pd_a\bar\c^2
		+\frac1{16}\c^2\bar\c^2\Box\c^2\Box\bar\c^2\big)\Big)\,.
\end{align}
where $w_a{}^b=\rmi\c\s^b\pd_a\bar\c$. 
The fermionic sector of the general $\cN=1$ vector self-dual model 
considered in \cite{KMcC} only differs by a rescaling of the last line above,
but only those with the fermionic sector given above can be mapped 
to the Akulov-Volkov action \cite{KMcC}.
The supersymmetry \eqref{eqn:BI-SUSY0} is projected to 
\begin{align} \label{eqn:BI-SUSY}
	\d_\h \c_\a &= \tfrac{1}\k \h_\a
	+ 2\k\h_\a\Big(\expt{\wb}-2\k^2(\expt{w}\expt\wb+\tfrac14\bar\c^2\Box\c^2)
	+4\k^4\big(\expt{w}^2\expt\wb+\expt{w}\expt\wb^2 \non\\
	&-\tfrac12\expt{\wb}\pd^a\c^2\pd_a\bar\c^2+\tfrac12\expt{w}\bar\c^2\Box\c^2
	+\tfrac14\expt\wb\c^2\Box\bar\c^2+\tfrac14\c^2\bar\c^2\Box\expt{\wb}\big) \\
	&-24\k^6\big(\expt{w}^2\expt{\wb}^2
	-\tfrac12\expt{w}\expt{\wb}\pd^a\c^2\pd_a\bar\c^2
	+\tfrac1{16}\c^2\bar\c^2\Box\c^2\Box\bar\c^2
	+\tfrac14(\expt{w}^2\bar\c^2\Box\c^2+\cc)\big)\Big) \non\\\non
	&+\rmi\k(\s^a\bar\h)_\a\pd_a\Big(\c^2\big(
	1-2\k^2\expt\wb+4\k^2\expt\wb^2+\k^4\bar\c^2\Box\c^2\big)\Big)\,.
\end{align}

By writing the action \eqref{eqn:BI} in the basis of Appendix \ref{sect:bases} 
and comparing to \eqref{eqn:GeneralAV}, 
we find the map that takes $S_\AV$ to $S_\BI$:
\begin{align} \label{eqn:AV2BI}
	\l_\a &= \c_\a-\k^2\big((1-\rmi\a_2^\rmi)\c_\a\expt{\wb} 
		+\tfrac\rmi2(\s^a\bar\c)_\a(\pd_a\c^2)\big)
	+\rmi{\k^4}(\s^a\bar\c)_\a(\pd_a\c^2)\expt{\b_7w +\b_8\wb} \\\non  					
	&+{\k^4}\c_\a\big(
	2(\rmi\a_2^\rmi+2\b_6+\b_8^*-3)\expt{w\wb}
	- (\tfrac\rmi2\a_2^\rmi+\b_4^*-4\b_6^\rmr+\b_7+\b_8^*-1)
		\expt{w}\expt{\wb} \\\non
	&+ (\tfrac32-2\rmi\a_2^\rmi-2\b_6)\expt{\wb^2} 
	+ \b_4 \expt{\wb}^2
	+ (\rmi\a_2^\rmi+2\b_6+\b_8^*-\tfrac52)\pd^a\c^2\pd_a\bar\c^2 
	+ \b_6\bar\c^2\Box\c^2\big) \\\non
	&+ {\k^6}\c_\a\big(
	  \expt{w\wb^2}
	- (2\a_2^\rmi(2\b_6^\rmi-\b_8^\rmi)-8\b_6^\rmr-2\b_7^\rmr-4\b_8^\rmr
		-\rmi\g_2^\rmi-2\g_5^\rmr+16)\expt{w\wb}\expt{\wb}\\\non
	&+ (\rmi\a_2^\rmi(2\b_6+\b_7+\b_8^*-3)+2\b_4^*+4\b_6^*
		+\b_7+4\b_8^\rmr-2\rmi\b_8^\rmi-18)\expt{w}\expt{\wb^2} \\\non
	&- (\tfrac12\a_2^\rmi(3\b_4^\rmi+4\b_6^\rmi+\b_7^\rmi+3\b_8^\rmi)
		-\b_4^\rmr-12\b_6^\rmr-\b_8^\rmr-\rmi\g_4^\rmi+45)
		\!\expt{w}\!\expt{\wb}^2\!
	+ \!\g_5\!\expt{\wb}\!\pd^a\c^2\pd_a\bar\c^2 \big)\,,
\end{align}
where we defined $8\b_6^\rmr=10-(\a_2^\rmi)^2$.
Once again, there are twelve free parameters \eqref{eqn:CoeffReIm} 
that correspond to the trivial symmetries of either action.

The pushforward of the AV supersymmetry using the map \eqref{eqn:AV2BI}
matches the supersymmetry \eqref{eqn:BI-SUSY} provided 
$\b_8^\rmr=\frac12=-\b_7^\rmr$ and all other free coefficients are zero.
In \cite{HK}, the theory of nonlinear realizations of supersymmetry
\cite{Ivanov2001}
was used to construct a scheme for finding the map from $S_\AV$ to $S_\BI$.
When explicitly carried out, this should reproduce \eqref{eqn:AV2BI} 
with the above choice of parameters.

\section{The chiral-scalar Goldstino action}\label{sect:CS}
In \cite{BG2} the $\cN=1$ tensor multiplet \cite{Siegel} was used to construct 
a Goldstone action for partial supersymmetry breaking. 
The tensor multiplet is described by a real linear scalar $L$ such that 
$D^2L=\Db^2L=0$. The authors of \cite{BG2} associated with $L$
the spinor superfield $\j_\a=\rmi D_\a L$, which, up to a switch in chirality, 
satisfies constraints and has a free action identical
to the field strength $W_\a$ used in the SBI action given above.
This correspondence allowed them to construct a Goldstone action by following 
the analogy with the SBI action. 
In \cite{RT} the same action was derived via a 
nilpotency constraint on the $\cN=2$ tensor multiplet.
The analogy with the SBI action is so close that the pure fermionic part of
the actions are exactly the same \cite{Thesis-McCarthy}, 
and thus there is no need to further examine it in this paper.

However, the tensor multiplet action can be dualised to obtain a Goldstone
action for partial supersymmetry breaking constructed from a chiral superfield.
The action obtained from this procedure is%
\footnote{We have rescaled relative to the conventions of Bagger and Galperin
in order to have an explicit dimensional coupling constant $\k$, 
a canonical fermion kinetic term  
and a canonical leading order Goldstino supersymmetry transformation.}
\begin{align} \label{eqn:BG-sf}
	S[\f,\bar\f] = \int  {\rm d}^4 x {\rm d}^4\q\, \cL(\f,\bar\f)\,,\quad
	\cL(\f,\bar\f) = 2\bar\f\f
		+\k^2(D^\a\f D_\a\f)(\Db_\da\bar\f\Db^\da\bar\f)f(A,B)\,,
\end{align}
where%
\footnote{Note that we use the opposite signature to that of \cite{BG2}.}
\begin{equation}\begin{gathered} \label{eqn:BG-f,A,B}
	f(A,B)^{-1} = 1 + \frac12A+\sqrt{1+A+B}\,,\quad
	A = 16\k^2\left(\pd_m\f\pd^m\bar\f-\frac1{16}D^2\f\Db^2\bar\f\right)\,,\\
	B = 2^6\k^4\left((\pd_m\f\pd^m\bar\f)^2
		-(\pd_m\f\pd^m\f)(\pd_n\bar\f\pd^n\bar\f)\right)\,.
\end{gathered}\end{equation}
In \cite{G-RPR} it was shown how the $D^2\f\Db^2\bar\f$ term may be removed
by a field redefinition of $\f$.  By a different field redefinition \cite{BG2}
it can also be shown that this action matches the leading order 
expression given in \cite{BG0}.
The action \eqref{eqn:BG-sf} is invariant under the nonlinear supersymmetry
transformation
\begin{align} \label{BG-sf-susy}
	\k\d_\h\f = \q\h + \frac{\k^2}4\h^\a\Db^2D_\a\cL\,.
\end{align}

Once again, projection to the fermion action is consistent with both 
the equations of motion \cite{Thesis-McCarthy} and the second supersymmetry.
We use the projection
\begin{align} \label{eqn:BG-proj}
	\f|=0\,,\quad D_\a\f|=\c_\a\,,\quad D^2\f|=0\,,
\end{align}
to obtain the fermionic action
\begin{align} \label{eqn:BG}
	S_\BG[\c,\bar\c] &= -\frac12\intx\Big(\expt{w+\wb}
	-2\k^2\big((\expt{w}^2+\expt{w^2}+\cc)+2\expt{w}\!\expt{\wb}
		+\tfrac12\pd^a\c^2\pd_a\bar\c^2\big) \non\\\non
	&+ 2\k^4\big(3(\expt{w^2}+3\expt{w}^2)\expt\wb+6\expt{w}\expt{w\wb}
		+2\expt{w^2\wb}-2\expt{w}\bar\c^2\Box\bar\c^2+\cc\big) \\
	&-8\k^6\big((\expt{w^2}\expt{\wb^2}+\cc)
		+\expt{w\wb w\wb}+4\expt{w^2\wb^2}+10\expt w^2\expt\wb^2 \\\non
	&\qquad+14\expt{w}\expt\wb\expt{w\wb}-\expt{w\wb}^2\big)\Big)\;,
\end{align}
and the supersymmetry transform 
\begin{align} \label{eqn:BG-SUSY}
	\d_\h\c_\a &= \frac1\k\h_\a  \non
	+ 2\rmi\k\big( (\h\s^a\bar\c)\pd_a\c_\a - (\h\c)(\pd\bar\c)_\a \big) \\\non
	&-2\k^3\Big( \c^2\big\{ (\h\s^a\bar\c)\pd_a(\pd\bar\c)_\a
			-              (\h\pd\bar\c)(\pd\bar\c)_\a  \big\}
		+ \bar\c^2\big\{ (\h\c)\Box\c_\a 
			+           (\h\s^a\pd\c)\pd_a\c_\a \big\} \\\non
	&\quad	- 2(\h\s^b\bar\c)(\c\s^a\pd_b\bar\c)\pd_a\c_\a
			+ 2(\h\c)(\bar\c\pd\c)(\pd\bar\c)_\a \Big) \\\non
	&+8\rmi\k^5\Big( 
		-\c^2\, \Big\{ \Big(\frac12\pd_a(\h\c)\pd^a\bar\c^2
			+ (\h\pd\bar\c)(\bar\c\pd\c) - (\h\s^b\bar\c)(\pd_b\bar\c\pd\c)
					\Big) \\ 
	&\quad	-(\h\s^b\bar\c)(\pd_b\bar\c\pd^a\bar\c)\pd_a\c_\a
		\Big\}(\pd\bar\c)_\a
		+ \bar\c^2 \Big\{ (\h\c)(\pd_b\c\s^b\pd\c)(\pd\bar\c)_\a \\\non
	&\quad	+	\big((\h\pd\bar\c)(\c\s^b\pd\c)
			+ \pd_a(\h\c)(\c\s^b\pd^a\bar\c) - (\h\c)(\pd_a\c\s^a\pd^b\bar\c) 
				\big)\pd_b\c_\a \Big\} \\\non
	&\quad	- \frac12\c^2\bar\c^2\Big\{
			  (\h\s^b\pd^a\bar\c)\pd_a\pd_b\c_\a + (\h\pd\bar\c)\Box\c_\a
			- \pd^a(\h\c)\pd_a(\pd\bar\c)_\a  + (\h\s^a\pd\c)\pd_a(\pd\bar\c)_\a
			  \Big\}	\Big)\allowdisplaybreaks\\\non
	&+ 8\k^7\c^2\bar\c^2\Big( \Big\{ 2\pd^a\pd^b\bar\c^2 \pd_b(\h\c)
			+ 5(\pd^a\bar\c\pd\c)(\h\pd\bar\c) 
			+ \frac12\pd_a(\h\c)\Box\bar\c^2  \Big\}\pd_a\c_\a		\\\non
	&\quad 	+ 5\Big\{(\pd_a\c\s^a\pd^b\bar\c)\pd_b(\h\c)
			- (\pd_a\c\s^a\pd\c)(\h\pd\bar\c)  \Big\}(\pd\bar\c)_\a
		\Big)\,.
\end{align}

The map that takes $S_\AV$ to $S_\BG$ is found to be
\begin{align} \label{eqn:AV2BG}
	\l_\a  &= \c_\a - \k^2\c_\a\expt{2w + (1-\rmi\a_2^\rmi)\wb} 
		+\frac\rmi2{\k^2}(\s^a\bar\c)_\a(\pd_a\c^2) 
	+\rmi{\k^4}(\s^a\bar\c)_\a(\pd_a\c^2)\expt{\b_7w +\b_8\wb} \non\\\non				
	&+{\k^4}\c_\a\big(
	 2(\left(2\b_6+\b_8^*+4\right)-\rmi\a_2^\rmi)\expt{w\wb}
	 - (\b_4^*-4\b_6^\rmr+\b_7+\b_8^*+\tfrac\rmi2\a_2^\rmi-14)
			\expt{w}\expt{\wb} \\\non
	&- (\tfrac{1}{2}+2\b_6)\expt{\wb^2} 
	+ \b_4 \expt{\wb}^2
	+ (2\b_6+\b_8^*+\tfrac52)\pd^a\c^2\pd_a\bar\c^2 
	+ \b_6\bar\c^2\Box\c^2\big) \\  
	&+ {\k^6}\c_\a\big(
	(4\rmi\a_2^\rmi-4\b_7-8\b_8^\rmr-11)\expt{w\wb^2} 
	+ \g_5\expt{\wb}\pd^a\c^2\pd_a\bar\c^2\\\non
	&- (2\a_2^\rmi(2\b_6^\rmi-\b_8^\rmi)+2\b_4^\rmr+24\b_6^\rmr+2\b_8^\rmr
		-\rmi\g_2^\rmi-2\g_5^\rmr+36)\expt{w\wb}\expt{\wb} \\\non
	&+ (\rmi\a_2^\rmi(\b_8^*+2\b_6+\b_7+4)+\b_7+2\b_8^\rmr+3)
		\expt{w}\expt{\wb^2} \\\non
	&+ (-\frac{1}{2}\a_2^\rmi(3\b_4^\rmi+4\b_6^\rmi+\b_7^\rmi+3\b_8^\rmi)
		-\b_4^\rmr+8\b_6^\rmr-\b_8^\rmr+\rmi\g_4^\rmi-9)\expt{w}\expt{\wb}^2
	 \big) \,,
\end{align}
where we have used $-8\b_6^\rmr=10+(\a_2^\rmi)^2$.

The pushforward of the AV supersymmetry using the map \eqref{eqn:AV2BG}
matches the supersymmetry \eqref{eqn:BG-SUSY} provided 
$\b_4^\rmr=5$, $\b_7^\rmr=\frac12$, $\b_8^\rmr=-\frac32$, $\g_5^\rmr=3$
and all other free coefficients are zero.

\section{Trivial symmetries and field redefinitions}\label{sect:Trivialities}
A trivial symmetry of a field theory is a symmetry transformation that reduces
to the identity transformation on-shell, i.e.\ 
\begin{align} \label{eqn:trivial}
	\vf^i \to \vf'^i = f^i(\vf,\dots) \xrightarrow{\text{on-shell}} \vf^i
	\quad \text{ such that} \quad S[\vf'] = S[\vf]\ .
\end{align}
It is well known (see, e.g., \cite{GT,HT}) 
that an infinitesimal symmetry transformation,
\begin{align} \label{eqn:infinitesimal_trivial}
	\vf^i \to \vf^i + \d\vf^i\,, 
\qquad S_{,i} [ \vf ] \,\d \vf^i =0\,, 
\end{align}
is trivial if and only if it can be written, 
using  DeWitt's condensed notation,  as
\begin{align} \label{eqn:infinitesimal_trivial2}
	\d\vf_i = S_{, j}[\vf ] \, \L^{ j i} [\vf ]\,, 
	\qquad \L^{ji} = -(-1)^{ij} \L^{ij}\,, 
\end{align}
for some super-antisymmetric matrix  $\L^{j i}$.
More generally, a transformation $\vf^i \to \vf'^i = f^i(\vf,\dots)$ 
is said to be trivial if it reduces to the identity map on the mass shell, 
\begin{align} \label{eqn:trivial_transform}
	\vf^i \to \vf'^i = f^i(\vf,\dots) \xrightarrow{\text{on-shell}} \vf^i\,.
\end{align}

The bulk of this section is dedicated to showing that the symmetries
of the KS action found in  section \ref{sect:KS} are all trivial.
We note that when two actions are related by a field redefinition, 
then trivial symmetries of one action are mapped onto trivial symmetries
of the other.  
Thus the triviality of the 12-parameter family of symmetries of $S_\KS$,
\eqref{eqn:KS2KS}, implies
the triviality of the same family of symmetries in any Goldstino action,%
\footnote{%
The free Majorana fermion action is not related to Goldstino action by the
field redefinitions \eqref{eqn:GeneralFieldRedef}. 
Apart from the four universal $O(\k^6)$ trivial symmetries,
it has only one other symmetry of the form \eqref{eqn:GeneralFieldRedef}. 
That symmetry is of $O(\k^4)$, trivial and, 
since there are no interaction terms, it has no higher order corrections terms.
} 
including the AV action.

The equations of motion that follow from \eqref{eqn:KS1} are
\begin{align} \label{eqn:EOM_KS}
	\rmi(\s_a\pd^a\bar\j)_\a 
	= \k^2\j_\a \Box\bar\j^2 (1-2\k^4\bar\j^2\Box\j^2)
	+ \k^6(\pd^a\j_\a)\j^2\pd_a(\bar\j^2\Box\bar\j^2)
\end{align}
and its complex conjugate. 
It's useful to contract the above with $\j^\a$ to get
\begin{align} \label{eqn:EOM_KS2}
	\expt{u} = \k^2\j^2 \Box\bar\j^2 (1-2\k^4\bar\j^2\Box\j^2)\,, \qquad
	\expt{\ub} = \k^2\bar\j^2 \Box\j^2 (1-2\k^4\j^2\Box\bar\j^2)\ .
\end{align}
We first apply these equations of motion to the general field redefinition
\eqref{eqn:GeneralFieldRedef} (with $\l\to\j$) to see when it is
trivial with respect to KS action. 
We can then specialise to the case of the symmetries of $S_\KS$.

Initially, we only use the contracted equation of motion \eqref{eqn:EOM_KS2}.
It is easy to see that this sends the terms associated with 
$\a_1$, $\b_2$, $\b_4$, $\b_7$, $\b_8$ and all $\g_{i\neq1}$ to zero.
While the $\a_2$ term becomes
\[ \j_\a\expt{\ub} \xrightarrow{\eqref{eqn:EOM_KS2}} 
	\k^2\j_\a\bar\j^2\Box\j^2\,, 
\]
and is thus mapped up to the $\b_6$ term. This leaves the field redefinition
\begin{align} \label{eqn:FieldRedefOS}
	\j \to \tilde\j_\a \xeq{\eqref{eqn:EOM_KS2}} 
	\j_\a&+\a_3\rmi{\k^2}(\s^a\bar\j)_\a(\pd_a\j^2) 
	+ \g_1{\k^6}\j_\a\expt{u\ub^2} \\\non
	&+{\k^4}\j_\a\big(\b_1\expt{u\ub}	+ \b_3\expt{\ub^2} 
		+ \b_5\pd^a\j^2\pd_a\bar\j^2 + (\a_2+\b_6)\bar\j^2\Box\j^2\big) \ .
\end{align}
Looking at the table of symmetries \eqref{12KS_Syms}, 
we see that combinations with $\b_1=2\b_5$ and $\b_3=-2(\a_2+\b_6)$ often occur.
Some spinor gymnastics shows that these combinations and no others
vanish on-shell
\begin{gather*} 
	\j_\a(2\expt{u\ub}+\pd^a\j^2\pd_a\bar\j^2)
	= \j_\a\left(2(\j\s^b\bar\j)(\pd_b\bar\j\tilde\s^a\pd_a\j)
		- \pd_a\j^2(\bar\j\tilde\s^a\s^b\pd_b\bar\j)\right)
	\xrightarrow{\eqref{eqn:EOM_KS}}	0\,, \\
	\j_\a\left(\expt{\ub^2}-\tfrac12\bar\j^2\Box\j^2\right)
	= \frac12\j_\a\bar\j^2\left(\pd^a\j\s_a+2\j\s_a\pd^a\right)\tilde\s_b\pd^b\j
	\xrightarrow{\eqref{eqn:EOM_KS}} - \k^2\j_\a\expt{u}\bar\j^2\Box\j^2
	\xrightarrow{\eqref{eqn:EOM_KS2}} 0 \,.
\end{gather*}

In summary, the general field redefinition \eqref{eqn:GeneralFieldRedef} 
(with $\l\to\j$) is trivial with respect to $S_\KS$
if the following conditions%
\footnote{The corresponding conditions for triviality of the field redefinition 
with respect to $S_\AV$ are a little more complicated:
\(\a_3=0\,,\; \b_1=2\b_5-\a_1-\a_2\,,\;
	\b_3=-(\a_2+2\b_6)\,,\; \g_1=\a_1+2\a_2-2(\b_5-2\b_6+\b_7+\b_8)\ .
\)
This can be used to directly show that the AV symmetries \eqref{eqn:AV2AV}
are trivial.
}  
on its coefficients hold
\begin{align} \label{eqn:KS_Trivial}
	\a_3=\g_1=0\,,\quad \b_1=2\b_5 \,,\quad \b_3=-2(\a_2+\b_6) \ .
\end{align}
These conditions specify the 24-parameter group 
of trivial transformations with respect to $S_\KS$.
It is now easy to check that all of the symmetries given by
\eqref{eqn:KS2KS} and \eqref{12KS_Syms} 
(and thus all Goldstino symmetries of the form \eqref{eqn:GeneralFieldRedef}) 
are trivial.

The above result can also be used to prove 
the triviality of any transformation relating 
the Ro\v{c}ek and the KS actions of sections \ref{sect:Rocek}
and \ref{sect:KS} respectively. 
Using the composition and inversion rules of Appendix \ref{sect:composition},
we find the set of maps that take $S_{\mathrm R}$ to $S_\KS$ can be 
parameterized as 
\begin{align} \label{eqn:RtoKS}
	\j_\a
	&\to \j_\a + \k^2\overbrace{\rmi \a_2^\rmi}^{\a_2}\j_\a\expt{\ub} 
		+\rmi{\k^2}(\s^a\bar\j)_\a(\pd_a\j^2)\big(\overbrace{0}^{\a_3}+				
		{\k^2}\expt{\b_7 u + \b_8 \ub}\big) \  \non\\\non						
	&+\k^4\j_\a\big(
	      \underbrace{2(2\rmi\a_2^\rmi+2\b_6+\b_8^*)}_{\b_1}\expt{u\ub}
		- (\b_4^*-4\b_6^\rmr+\b_7+\b_8^*)\expt{u}\expt{\ub} \\\non
		&\quad- 2\underbrace{(\rmi\a_2^\rmi+\b_6)}_{\b_3}\expt{\ub^2} 
		+ \b_4 \expt{\ub}^2
		+ \underbrace{(2\rmi\a_2^\rmi + 2\b_6 + \b_8^*)}_{\b_5}
			\pd^a\j^2\pd_a\bar\j^2 
		+ \b_6\bar\j^2\Box\j^2\big) ~~~~\\\non
	&+ {\k^6}\j_\a\big(\underbrace{0}_{\g_1}\expt{u\ub^2}
		+ (\rmi\g_2^\rmi - 2\a_2^\rmi(2\b_6^\rmi-\b_8^\rmi)
			+2\g_5^\rmr)\expt{u\ub}\expt{\ub} 
		+ \g_5\expt{\ub}\pd^a\j^2\pd_a\bar\j^2\\
		&\quad+ (\rmi\a_2^\rmi(2\b_6+\b_7+\b_8^*-4) 
			+ 2(6\b_6^\rmr-2\rmi\b_6^\rmi+\b_4^*+\b_8^*))\expt{u}\expt{\ub^2} \\
		&\quad+ (\rmi\g_4^\rmi 
			- \tfrac12\a_2^\rmi(3\b_4^\rmi+4\b_6^\rmi+\b_7^\rmi+3\b_8^\rmi)
			+2\b_4^\rmr+12\b_6^\rmr+2\b_8^\rmr)\expt{u}\expt{\ub}^2
		 \big) \,,~~~~ \non
\end{align}
where $-8\b_6^\rmr=(\a_2^\rmi)^2$.
This directly demonstrates the off-shell equivalence of the models.
It is then easy to check that the conditions of \eqref{eqn:KS_Trivial} are
satisfied, and so \eqref{eqn:RtoKS} reduces to the identity map when on-shell
with respect to $S_\KS$.

\section{Conclusion}\label{sect:Conc}
In this paper we have presented the most general field redefinition 
\eqref{eqn:GeneralFieldRedef}
that can be used to relate any Goldstino model to the AV action.
This has been used to prove,
by explicit construction of the field redefinitions, 
the equivalence between most Goldstino models that occur in the literature.  
Each of these field redefinitions has twelve free parameters that follow 
from the trivial symmetries possessed by all Goldstino models.
These trivial symmetries preserve the form of the Goldstino actions, 
but not the form of the associated nonlinear supersymmetry transformations,
which implies that each Goldstino model has a 12 parameter family of 
on-shell equivalent nonlinear supersymmetry transformations.
Except for the KS action, all Goldstino models studied in this paper
have a natural nonlinear supersymmetry that acts on the Goldstino. 
We have shown that all of these natural supersymmetry transformations 
may be obtained from the pushforward of the AV supersymmetry 
for a particular choice of the trivial symmetry parameters.
It is interesting to note that in all of the models studied in this paper,
the above choice of trivial symmetry parameters always leads to purely
real coefficients in the field redefinition \eqref{eqn:GeneralFieldRedef}.
For the KS Goldstino, as there is no known natural nonlinear supersymmetry,
we simply set all trivial symmetry parameters to zero and gave the resulting
nonlinear supersymmetry.

Most other proofs of the equivalence between the various Goldstino models 
which have appeared in the literature,  
are based on exploiting the explicit structure of nonlinear supersymmetry 
available on either side of the equivalence.
Ro\v{c}ek \cite{Rocek1978} obtained the first result by directly looking for a 
map that took the linear supersymmetry of his constrained superfield 
to that of Akulov and Volkov. 
Using their general theory of spontaneously broken  supersymmetry 
\cite{Ivanov1977,Ivanov1978,Ivanov1982},  Ivanov and Kapustnikov
constructed homogeneously transforming nonlinear superfields that linked 
the two models upon  removal of the spectator fields. 
Finally, Samuel and Wess \cite{SamuelWess1983} used the nonlinear supersymmetry 
to construct a linear superfield \eqref{eqn:chAV_Superfield}
and showed how the constraints underlying both the Ro\v{c}ek \cite{Rocek1978}
and the Lindstr\"om-Ro\v{c}ek \cite{LR} models 
can be solved using this superfield.

The KS model \cite{KS} is the stand-out case since, 
after elimination of the auxiliary field, 
there is no obvious off-shell realisation of supersymmetry
acting on the component field $\j_\a$ 
\cite{Casalbuoni1989}. 
Thus the identification of the Goldstino and the application of methods 
based on nonlinear representation theory becomes problematic.
In \cite{Luo2009}, this problem was overcome by noticing that there is a 
simple composite Goldstino that can be constructed out of the components 
\eqref{eqn:ComponentProjections-Rocek} 
of the constrained chiral superfield $\F^2=0$. 
Their argument goes that for an unconstrained chiral superfield one defines 
$\tilde\j_\a=(\sqrt2\k F)^{-1}\j_\a$
and uses \eqref{eqn:FreeSusy} to calculate its supersymmetry transformation 
\begin{align*} 
	\d_\x\tilde\j_\a 
	= \frac{\x_\a}\k-2\rmi\k(\tilde\j\s^a\bar\x)\pd_a\tilde\j_\a
		+\frac{\rmi}{\k F}(\s^a\bar\x)_\a\pd_a\Big(\f-\frac{\j^2}{2F}\Big)\,.
\end{align*}
This means that if one constrains the superfield so that 
$\f-\frac{\j^2}{2F}=0 \iff\F^2=0$ then the field $\tilde\j_\a$ 
transforms exactly as the chiral AV Goldstino \eqref{eqn:chAV_SUSY}.
In \cite{Liu2010} this observation was used to apply the tools of the nonlinear
realisation of supersymmetry \cite{Ivanov1978,Ivanov1982}  
to prove the equivalence of the KS, AV and chiral AV models by deriving
them all from the same superfield action.
However, their methods did not yield an explicit mapping from $S_\AV[\l]$
to $S_\KS[\j]$ and thus could not give the off-shell supersymmetry
of the KS Goldstino $\j_\a$, both are problems solved in this paper.

In conclusion, we should also mention another approach 
to constructing a field redefinition from the KS to the AV
action which was advocated  in 
\cite{Zheltukhin2010,Zheltukhin2010b,Zheltukhin2010c}.
As shown in \cite{letter}, the first two papers in this series contained
a small oversight that rendered the results incorrect. This oversight
was corrected
\footnote{The location of the error and the method of correction were pointed 
out to Professor Zheltukhin in an email from SJT dated 16th September 2010.} 
in (the proofs for) the published version of \cite{Zheltukhin2010}
and a correct, lowest order, field redefinition was given in 
\cite{Zheltukhin2010c}.
This field redefinition matches the simple one given in 
equation (8) of \cite{letter} but fails to find the 
one trivial symmetry that exists at that order.
\\

\noindent
{\bf Acknowledgements:}\\
SMK thanks Martin Ro\v{c}ek for discussions and useful comments on the manuscript.
SJT is grateful to Ian McArthur and Paul Abbott for useful discussions and
to Gabriele Tartaglino-Mazzucchelli for constructive comments on the manuscript.
SJT is also indebted to Robert Stamps for the account on the machine where
some of the larger calculations were performed.
The work of SMK is supported in part by the Australian Research Council.

\appendix
\section{Minimal basis for Goldstino actions}\label{sect:bases}
In order to compare different Goldstino actions, we need to be able
to write all terms in a common basis of Lorentz invariant terms.
We restrict our attention to terms that occur in Goldstino actions and
thus have the structure given in \eqref{eqn:GoldStruct}.
Obviously, there is a lot of freedom in the choice of such a basis. 
We have chosen a basis where as many elements as possible can be 
written as traces of the matrices $v=(v_a{}^b)$ and $ \vb = (\vb_a{}^b)$
defined in  \eqref{defn:Xi v vb}.

We choose the minimal basis for 4-fermion terms to be
\begin{align} \label{eqn:4fermion_basis}
	\expt{v^2},	\quad	\expt{\vb^2},	\quad 	\expt{v}\expt{\vb}, \quad
	\expt{v}^2,	\quad	\expt{\vb}^2,	\quad	\pd^a\l^2\pd_a\bar\l^2\,.
\end{align}
In this basis, the structure that occurs in the 
AV action \eqref{eqn:AV2} becomes
\begin{align}\label{<v><vb>-<vvb>} 
	2\left(\expt{v}\expt{\vb}-\expt{v\vb}\right) 
	&= \expt{v}^2-\expt{v^2}+\expt{\vb}^2-\expt{\vb^2}\ .
\end{align}

The 6-fermion basis is chosen to be
\begin{equation} \label{eqn:6fermion_basis}
\begin{gathered}
	\expt{v^2\vb}, \;
	\expt{v}\expt{v\vb},\;				\expt{\vb}\expt{v\vb},\;
	\expt{v^2}\expt{\vb},\; 			\expt{v}\expt{\vb^2},\;
	\expt{v}^2\expt{\vb},\; 			\expt{v}\expt{\vb}^2, \\
	\expt{v}\pd^a\l^2\pd_a\bar\l^2,\;	\expt{\vb}\pd^a\l^2\pd_a\bar\l^2,\;
	\rmi\l\s^a\bar\l(\expt{v}\oLRa\pd_a\expt{\vb})\ ,
\end{gathered}
\end{equation}
where the first term is the only one that is neither real 
nor in a complex conjugate pair 
and in the last term we use the symbol 
\( x\oLRa\pd_ay = x(\pd_ay)-(\pd_ax)y \ .\)
When writing 6-fermion expressions, we will often use the overcomplete
basis that includes the complex conjugate of the first term
\begin{align} 
	\expt{v\vb^2} = \expt{v^2\vb} + \half\Big(\expt{v}^2\expt{\vb}
		-2\expt{v}\expt{v\vb}-\expt{v^2}\expt{\vb}	-\cc \Big)\ .
\end{align}
Most actions in this paper have also been simplified
by rewriting the last basis element in \eqref{eqn:6fermion_basis} 
using the extra terms 
$\expt{v}\bar\l^2\Box\l^2$ and $\expt{\vb}\l^2\Box\bar\l^2$:
\begin{align} 
	\rmi\l\s^a\bar\l(\expt{v}\oLRa\pd_a\expt{\vb})
	= \frac12\expt{v^2}\expt{\vb}-\expt{v}\expt{v\vb}
		-\frac14\expt{v}\bar\l^2\Box\l^2
		- \frac12\expt{v}\pd^a\l^2\pd_a\bar\l^2
		+ \cc
\end{align}

All 8-fermion terms can be written as traces. 
We choose the basis
\begin{equation}
\begin{gathered}
	\expt{v}\expt{v\vb^2},\quad 			\expt{\vb}\expt{v^2\vb},\quad
	\expt{v}^2\expt{\vb^2},\quad\;			\expt{v^2}\expt{\vb}^2, \\\;
	\expt{v^2}\expt{\vb^2},\quad			\expt{(v\vb)^2},\quad
	\expt{v}\expt{\vb}\expt{v\vb},\quad		\expt{v}^2\expt{\vb}^2 \ .
\end{gathered}
\end{equation}
The identities needed to show the vanishing of the $O(\k^6)$ 
terms in $S_\AV$ are 
\begin{align}  \nonumber
	\expt{v^2\vb^2} &=
		\expt{v}\expt{v\vb^2}+\expt{\vb}\expt{v^2\vb}
		-\expt{(v\vb)^2} - \expt{v}\expt{\vb}\expt{v \vb}
		+\half\expt{v}^2\expt{\vb}^2+\half\expt{v^2}\expt{\vb^2}\,, \\
	2\expt{v\vb}^2 &=
		\expt{v}^2\expt{\vb^2}+\expt{v^2}\expt{\vb}^2
		+\expt{v^2}\expt{\vb^2}
		-2\expt{(v\vb)^2}+\expt{v}^2\expt{\vb}^2\ .
\end{align}
The 8-fermion term that occurs in the KS action is
\begin{align} 
	\frac14\l^2\bar\l^2\Box\l^2\Box\bar\l^2 = 
	\expt{v}^2\expt{\vb^2} + \expt{v^2}\expt{\vb}^2 +
	\expt{v^2}\expt{\vb^2} + \expt{v}^2\expt{\vb}^2 \ .
\end{align}

The proof that the above basis is minimal is based on computer calculation
and is contained in the attached Mathematica program.

\section{Composition rule for field redefinitions}\label{sect:composition}
By direct calculation, 
the composition of a transformation \eqref{eqn:GeneralFieldRedef}
with coefficients $x_i$, $y_j$ \& $z_k$ followed by one with
coefficients $a_i$, $b_j$ \& $c_k$ 
is shown to be the same as the transformation with coefficients
$\a_i$, $\b_j$ and $\g_k$ where
\begin{subequations}\label{eqn:comp}
\begin{align}\label{eqn:comp_alpha}
  \a_1 &= a_ 1+x_ 1\,, \quad
  \a_2 = a_ 2+x_ 2\,, \quad
  \a_3 = a_ 3+x_ 3 
  \allowdisplaybreaks\\ 
  \b_1 &= b_ 1+y_ 1+a_ 2 x_ 2+4 a_ 2 x_ 3+4 a_ 3 x_ 3-x_ 1 a_ 1^*-2 x_ 1 a_ 3^*-2 x_ 2 a_ 3^*-4 x_ 3 a_ 3^* \\  
  \b_2 &= b_ 2+y_2 +2 a_ 2 x_ 1+2 a_ 3 x_ 1+2 a_ 1 x_ 2+a_ 2 x_ 2+2 a_ 3 x_ 2+4 a_ 2 x_ 3+4 a_ 3 x_ 3+x_ 2 a_ 2^* \allowdisplaybreaks \\ 
  \b_3 &= b_ 3+y_ 3-\tfrac{a_ 2 x_ 2}{2}-2 a_ 2 x_ 3-2 a_ 3 x_ 3+\tfrac{x_ 1 a_ 1^*}{2} \\  
  \label{eqn:comp_beta} 
  \b_4 &= b_ 4+y_ 4+\tfrac{3 a_ 2 x_ 2}{2}-2 a_ 2 x_ 3-2 a_ 3 x_ 3+\tfrac{x_ 1 a_ 1^*}{2}+x_ 2 a_ 1^*  \allowdisplaybreaks \\ 
  \b_5 &= b_ 5+y_ 5+\tfrac{a_ 2 x_ 2}{2}+a_ 3 x_ 2+2 a_ 2 x_ 3+2 a_ 3 x_ 3-\tfrac{x_ 1 a_ 1^*}{2}-x_ 2 a_ 3^* \\ 
  \b_6 &= b_ 6+y_ 6+\tfrac{a_ 2 x_ 2}{4}+a_ 3 x_ 2+a_ 2 x_ 3+a_ 3 x_ 3-\tfrac{x_ 1 a_ 1^*}{4} \allowdisplaybreaks \\ 
  \b_7 &= b_ 7+y_ 7-\tfrac{a_ 1 x_ 2}{2}-\tfrac{a_ 2 x_ 2}{2}-a_ 3 x_ 2-2 a_ 2 x_ 3-2 a_ 3 x_ 3+\tfrac{x_ 1 a_ 1^*}{2}+\tfrac{x_ 1 a_ 2^*}{2} \\\non
  &\quad+x_ 3 a_ 2^*+x_1 a_ 3^*+x_2 a_ 3^*+2 x_ 3 a_ 3^* \\ 
  \b_8 &= b_ 8+y_ 8+a_ 3 x_ 2+2 a_ 2 x_ 3+2 a_ 3 x_ 3+x_ 3 a_ 1^* \allowdisplaybreaks \\ 
  \g_1 &= c_ 1+z_ 1-4 x_ 2 a_ 3^2+2 y_ 1 a_ 3-4 y_ 3 a_ 3
  -4 y_ 5 a_ 3-4 y_ 8 a_ 3-b_ 1 x_ 2+2 b_ 5 x_ 2 \\ \non
  &\quad-2 y_ 1 a_ 3^*-4 y_ 7 a_ 3^*+x_1 b_ 1^*+2 x_ 3 b_ 1^*
  -2 x_ 1 b_ 5^*-4 x_ 3 b_ 5^* \allowdisplaybreaks \\ 
  \g_2 &= c_ 2+z_ 2-4 x_ 2 a_ 3^2-2 a_ 2 x_ 2 a_ 3+4 y_ 1 a_ 3
  -4 y_ 4 a_ 3-4 y_ 5 a_ 3+4 y_ 7 a_ 3-4 y_ 8 a_ 3\\ \non
  &\quad-2 x_ 2 a_ 1^* a_ 3+2 x_ 1 a_ 2^* a_ 3+2 x_ 2 a_ 2^* a_ 3+4 x_ 3 a_ 2^* a_ 3+4 x_ 2 a_ 3^* a_ 3-a_ 1 a_ 2 x_ 2+b_ 1 x_ 2-b_ 2 x_ 2\\ \non
  &\quad+2 b_ 5 x_ 2+2 b_ 7 x_ 2-2 b_ 8 x_ 2+3 a_ 2 y_ 1+2 a_ 1 y_ 3-2 a_ 1 y_ 4+6 a_ 2 y_ 7-a_ 1 x_ 2 a_ 1^*+2 y_ 1 a_ 1^*\\ \non
  &\quad+4 y_ 7 a_ 1^*+2 a_ 2 x_ 1 a_ 2^*+4 a_ 2 x_ 3 a_ 2^*-y_1 a_ 2^*+y_2 a_ 2^*+2 y_ 7 a_ 2^*+4 y_ 8 a_ 2^*-2 y_ 1 a_ 3^*+4 y_ 8 a_ 3^*\\ \non
  &\quad+x_1 b_ 1^*+x_2 b_ 1^*+2 x_ 3 b_ 1^*+x_1 b_ 2^*+2 x_ 3 b_ 2^*
  -2 x_ 1 b_ 5^*-4 x_ 3 b_ 5^*  \allowdisplaybreaks \\ 
  \g_3 &=c_ 3+z_ 3 + 2 x_ 1 a_ 3^2+2 x_ 2 a_ 3^2+4 x_ 3 a_ 3^2+3 a_ 1 x_ 2 a_ 3+2 y_ 2 a_ 3+2 y_ 7 a_ 3+2 y_ 8 a_ 3\\ \non
  &\quad+2 x_ 2 a_ 2^* a_ 3+2 x_ 2 a_ 3^* a_ 3+2 b_ 3 x_ 1+4 b_ 6 x_ 1+\tfrac{b_ 1 x_ 2}{2}+b_ 3 x_ 2+2 b_ 6 x_ 2+b_ 7 x_ 2+4 b_ 3 x_ 3\\ \non
  &\quad+8 b_ 6 x_ 3+2 a_ 1 y_ 3+2 a_ 1 y_ 6+a_ 1 y_ 8+\tfrac{y_ 1 a_ 2^*}{2}+2 y_ 3 a_ 2^*-2 y_ 5 a_ 2^*+4 y_ 6 a_ 2^*-y_7 a_ 2^*+y_1 a_ 3^* \allowdisplaybreaks \\ \non
  &\quad+2 y_ 3 a_ 3^*-2 y_ 5 a_ 3^*+4 y_ 6 a_ 3^*-\tfrac{x_ 1 b_ 1^*}{2}-x_ 3 b_ 1^* \\ 
  \g_4 &=c_ 4+z_ 4+ x_ 1 a_2^2+\tfrac{1}{2} x_ 2 a_2^2+2 x_ 3 a_2^2+2 a_ 3 x_ 1 a_ 2+\tfrac{3}{2} a_ 1 x_ 2 a_ 2+2 a_ 3 x_ 2 a_ 2+4 a_ 3 x_ 3 a_ 2\\ \non
  &\quad+3 y_ 2 a_ 2-y_ 3 a_ 2+y_ 4 a_ 2+3 y_ 7 a_ 2+3 y_ 8 a_ 2+x_ 1 a_ 1^* a_ 2+\tfrac{1}{2} x_ 2 a_ 1^* a_ 2+2 x_ 3 a_ 1^* a_ 2+x_ 1 a_ 2^* a_ 2\\ \non
  &\quad+2 x_ 2 a_ 2^* a_ 2+2 x_ 3 a_ 2^* a_ 2+2 x_ 1 a_ 3^* a_ 2+2 x_ 2 a_ 3^* a_ 2+4 x_ 3 a_ 3^* a_ 2+2 a_ 3^2 x_ 1+2 b_ 4 x_ 1+4 b_ 6 x_ 1\\ \non
  &\quad+2 b_ 8 x_ 1+2 a_ 3^2 x_ 2+3 a_ 1 a_ 3 x_ 2+\tfrac{3 b_ 2 x_ 2}{2}+b_ 4 x_ 2+2 b_ 6 x_ 2+2 b_ 7 x_ 2+b_ 8 x_ 2+4 a_ 3^2 x_ 3+4 b_ 4 x_ 3\\ \non
  &\quad+8 b_ 6 x_ 3+4 b_ 8 x_ 3+4 a_ 3 y_ 2-2 a_ 3 y_ 3+2 a_ 1 y_ 4+2 a_ 3 y_ 4+2 a_ 1 y_ 6+4 a_ 3 y_ 7+a_ 1 y_ 8+4 a_ 3 y_ 8\\ \non
  &\quad+a_ 3 x_ 1 a_ 1^*+\tfrac{3}{2} a_ 1 x_ 2 a_ 1^*+a_3 x_ 2 a_ 1^*+2 a_ 3 x_ 3 a_ 1^*+\tfrac{y_ 1 a_ 1^*}{2}+\tfrac{3 y_ 2 a_ 1^*}{2}+y_ 7 a_ 1^*+a_3 x_ 1 a_ 2^*+3 a_ 3 x_ 2 a_ 2^*\\ \non
  &\quad+2 a_ 3 x_ 3 a_ 2^*+\tfrac{y_ 2 a_ 2^*}{2}+2 y_ 4 a_ 2^*-2 y_ 5 a_ 2^*+4 y_ 6 a_ 2^*+2 y_ 8 a_ 2^*+2 a_ 3 x_ 1 a_ 3^*+2 a_ 3 x_ 2 a_ 3^*+4 a_ 3 x_ 3 a_ 3^*\\ \non
  &\quad+y_2 a_ 3^*+2 y_ 4 a_ 3^*-2 y_ 5 a_ 3^*+4 y_ 6 a_ 3^*+2 y_ 8 a_ 3^*+\tfrac{x_ 1 b_ 2^*}{2}+x_ 2 b_ 2^*+x_3 b_ 2^*+x_1 b_ 7^*+x_2 b_ 7^*+2 x_ 3 b_ 7^* \allowdisplaybreaks \\ 
  \g_5 &= c_ 5+z_ 5-\tfrac{a_ 1 a_ 2 x_ 2}{2} +a_ 2 a_ 3 x_ 2-\tfrac{b_ 1 x_ 2}{2}-\tfrac{b_ 2 x_ 2}{2}+2 b_ 5 x_ 2+b_ 7 x_ 2-b_ 8 x_ 2+a_ 1 y_ 3-a_ 2 y_ 3\\ \non
  &\quad-2 a_ 3 y_ 3-a_ 1 y_ 4+3 a_ 2 y_ 5+2 a_ 3 y_ 5-4 a_ 2 y_ 6-4 a_ 3 y_ 6+3 a_ 2 y_ 7+2 a_ 3 y_ 7+a_ 2 y_ 8-\tfrac{1}{2} a_ 1 x_ 2 a_ 1^*\\ \non
  &\quad+\tfrac{y_ 1 a_ 1^*}{2}+y_ 7 a_ 1^*+a_2 x_ 1 a_ 2^*+a_3 x_ 1 a_ 2^*+a_3 x_ 2 a_ 2^*+2 a_ 2 x_ 3 a_ 2^*+2 a_ 3 x_ 3 a_ 2^*-\tfrac{y_ 1 a_ 2^*}{2}+\tfrac{y_ 2 a_ 2^*}{2}\\ \non
  &\quad+y_ 7 a_ 2^*+2 y_ 8 a_ 2^*+2 a_ 2 x_ 1 a_ 3^*+2 a_ 3 x_ 1 a_ 3^*+4 a_ 2 x_ 3 a_ 3^*+4 a_ 3 x_ 3 a_ 3^*-y_1 a_ 3^*+y_2 a_ 3^*+2 y_ 8 a_ 3^*\\ \non
  &\quad+\tfrac{x_ 1 b_ 1^*}{2}+x_ 3 b_ 1^*+\tfrac{x_ 1 b_ 2^*}{2}+x_ 3 b_ 2^*+x_2 b_ 5^*+x_1 b_ 7^*+2 x_ 3 b_ 7^* \ . 
\end{align}
\end{subequations}
The composition of field redefinitions is the group law for the noncommutative 
group $G$ of all transformations that relate all of the Goldstino models.
When the field redefinitions are restricted to the trivial symmetries 
of any particular Goldstino action, then we obtain the group law 
for the subgroup $H$ of such trivial symmetry transformations. 

The composition rule \eqref{eqn:comp_alpha} with $\a_i=\b_j=\g_k=0$ 
can be solved to give the inversion rule for field redefinitions.
The inverse of a field redefinition with coefficients $a_i$, $b_j$ \& $c_k$
is one with coefficients $x_i$, $y_j$ \& $z_k$, where
{\allowdisplaybreaks
\begin{subequations}\label{eqn:inversion}
\begin{align} \label{eqn:inversion_alpha}
  a_1+x_1 &= 0 \,,\quad
  a_2+x_2 = 0 \,,\quad
  a_3+x_3 = 0 \\ 
  b_1+y_1 &= -|a_1|^2-4 |a_3|^2+a_2^2+4 a_3^2+4 a_2 a_3-2 a_1 a_3^*-2 a_2 a_3^* \\
  b_2+y_2 &= |a_2|^2+a_2^2+4 a_3^2+4 a_1 a_2+2 a_1 a_3+6 a_2 a_3\,,\\
  b_3+y_3 &= \tfrac{|a_1|^2}{2}-\tfrac{a_2^2}{2}-2 a_3^2-2 a_2 a_3 \,,\\
  b_4+y_4 &= \tfrac{|a_1|^2}{2}+\tfrac{3 a_2^2}{2}-2 a_3^2-2 a_2 a_3+a_2 a_1^* \\
  b_5+y_5 &= -\tfrac{|a_1|^2}{2}+\tfrac{a_2^2}{2}+2 a_3^2+3 a_2 a_3-a_2 a_3^* \,,\\
  b_6+y_6 &= -\tfrac{|a_1|^2}{4}+\tfrac{a_2^2}{4}+a_3^2+2 a_2 a_3 \\
  b_7+y_7 &= \tfrac{|a_1|^2}{2}+2 |a_3|^2-\tfrac{a_2^2}{2}-2 a_3^2-\tfrac{a_1 a_2}{2}-3 a_2 a_3+\tfrac{a_1 a_2^*}{2}+a_3 a_2^*+a_1 a_3^*+a_2 a_3^*  \\ 
  b_8+y_8 &= 2 a_3^2+3 a_2 a_3+a_1^* a_3  \\ 
  c_1+z_1 &= 2 a_3 |a_1|^2+4 a_2 a_3^2+4 |a_3|^2 a_1-4 |a_3|^2 a_2+8 |a_3|^2 a_3-2 a_2^2 a_3-a_2 b_1 \\ \non
  &+2 a_3 b_1-4 a_3 b_3+2 a_2 b_5-4 a_3 b_5-4 a_3 b_8+4 a_3^2 a_1^*+4 |a_3|^2 a_2^*-2 a_1 a_2 a_3^* \\ \non
  &-2 b_1 a_3^*-4 b_7 a_3^*+2 a_1 a_2^* a_3^*+a_1 b_1^*+2 a_3 b_1^*-2 a_1 b_5^*-4 a_3 b_5^* \\
  c_2+z_2 &= 3 a_2 |a_1|^2+2 a_3 |a_1|^2-4 a_2^* |a_1|^2-2 a_3^* |a_1|^2+6 a_1 a_2^2+8 a_2 a_3^2-a_1 a_2^*{}^2 \\ \non
  &-a_2 a_2^*{}^2-2 a_3 a_2^*{}^2-4 a_1 a_3^*{}^2-4 a_2 a_3^*{}^2-8 a_3 a_3^*{}^2-4 |a_2|^2 a_1+4 |a_3|^2 a_1+|a_2|^2 a_2 \\ \non
  &-8 |a_2|^2 a_3+8 |a_3|^2 a_3+10 a_2^2 a_3+2 a_1 a_2 a_3+4 a_2 b_1+4 a_3 b_1-a_2 b_2+2 a_1 b_3-2 a_1 b_4\\ \non
  &-4 a_3 b_4+2 a_2 b_5-4 a_3 b_5+8 a_2 b_7+4 a_3 b_7-2 a_2 b_8-4 a_3 b_8-4 |a_3|^2 a_1^*+4 a_3^2 a_1^*\\ \non
  &+6 a_2 a_3 a_1^*+2 b_1 a_1^*+4 b_7 a_1^*-8 |a_3|^2 a_2^*-4 a_3^2 a_2^*-2 a_1 a_3 a_2^*-b_1 a_2^*+b_2 a_2^*+2 b_7 a_2^*\\ \non
  &+4 b_8 a_2^*-8 a_3 a_1^* a_2^*-4 |a_2|^2 a_3^*+2 a_2^2 a_3^*-2 b_1 a_3^*+4 b_8 a_3^*-4 a_1 a_2^* a_3^*+a_1 b_1^*+a_2 b_1^* \\ \non
  &+2 a_3 b_1^*+a_1 b_2^*+2 a_3 b_2^*-2 a_1 b_5^*-4 a_3 b_5^* \\
  c_3+z_3 &= -4 (a_3)^3-2 a_1 a_3^2-10 a_2 a_3^2-2 a_1^* a_3^2-2 a_2^* a_3^2-2 |a_1|^2 a_3-3 |a_2|^2 a_3 \\\non
  &-4 |a_3|^2 a_3-a_2^2 a_3+a_2^*{}^2 a_3+4 a_3^*{}^2 a_3-7 a_1 a_2 a_3+2 b_2 a_3+4 b_3 a_3+8 b_6 a_3+2 b_7 a_3\\\non
  &+2 b_8 a_3-a_1 a_2^* a_3-b_1^* a_3+\tfrac{1}{2} a_1 a_2^2+\tfrac{1}{2} a_1 a_2^*{}^2+2 a_1 a_3^*{}^2-\tfrac{|a_1|^2 a_1}{2}-\tfrac{|a_2|^2 a_1}{2}-2 |a_3|^2 a_1x\\\non
  &+2 a_2^2 a_3^*+\tfrac{a_2 b_1}{2}+4 a_1 b_3+a_2 b_3+6 a_1 b_6+2 a_2 b_6+a_2 b_7+a_1 b_8+4 |a_3|^2 a_2^*+\tfrac{b_1 a_2^*}{2}\\\non
  &+2 b_3 a_2^*-2 b_5 a_2^*+4 b_6 a_2^*-b_7 a_2^*+b_1 a_3^*+2 b_3 a_3^*-2 b_5 a_3^*+4 b_6 a_3^*+2 a_1 a_2^* a_3^*-\tfrac{a_1 b_1^*}{2} \\
  c_4+z_4 &= -3 (a_2)^3-\tfrac{23}{2} a_1 a_2^2-20 a_3 a_2^2-2 a_1^* a_2^2-5 a_3^* a_2^2-\tfrac{13 |a_1|^2 a_2}{2}-\tfrac{9 |a_2|^2 a_2}{2}\\\non
  &-14 |a_3|^2 a_2-30 a_3^2 a_2-\tfrac{1}{2} a_2^*{}^2 a_2-2 a_3^*{}^2 a_2-18 a_1 a_3 a_2+\tfrac{9 b_2 a_2}{2}-b_3 a_2+2 b_4 a_2\\\non
  &+2 b_6 a_2+5 b_7 a_2+4 b_8 a_2-10 a_3 a_1^* a_2-5 a_1 a_3^* a_2-2 a_1^* a_3^* a_2+b_2^* a_2+b_7^* a_2-12 (a_3)^3\\\non
  &-6 a_1 a_3^2-\tfrac{|a_1|^2 a_1}{2}-\tfrac{5}{2}|a_2|^2 a_1-4 |a_3|^2 a_1-5 |a_1|^2 a_3-9 |a_2|^2 a_3-8 |a_3|^2 a_3+4 a_3 b_2\\\non
  &-2 a_3 b_3+4 a_1 b_4+6 a_3 b_4+6 a_1 b_6+8 a_3 b_6+4 a_3 b_7+3 a_1 b_8+8 a_3 b_8-\tfrac{7 |a_2|^2 a_1^*}{2}\\\non
  &-2 |a_3|^2 a_1^*-8 a_3^2 a_1^*+\tfrac{b_1 a_1^*}{2}+\tfrac{3 b_2 a_1^*}{2}+b_7 a_1^*-\tfrac{3 |a_1|^2 a_2^*}{2}-4 a_3^2 a_2^*-2 a_1 a_3 a_2^*+\tfrac{b_2 a_2^*}{2}+2 b_4 a_2^*\\\non
  &-2 b_5 a_2^*+4 b_6 a_2^*+2 b_8 a_2^*-3 a_3 a_1^* a_2^*-|a_1|^2 a_3^*-3 |a_2|^2 a_3^*+b_2 a_3^*+2 b_4 a_3^*-2 b_5 a_3^*\\\non
  &+4 b_6 a_3^*+2 b_8 a_3^*+\tfrac{a_1 b_2^*}{2}+a_3 b_2^*+a_1 b_7^*+2 a_3 b_7^* \\
  c_5+z_5 &= \tfrac{(a_2)^3}{2}+3 a_1 a_2^2+4 a_3 a_2^2+\tfrac{|a_1|^2 a_2}{2}+\tfrac{|a_2|^2 a_2}{2}-10 |a_3|^2 a_2+4 a_3^2 a_2-\tfrac{1}{2} a_2^*{}^2 a_2\\\non
  &-2 a_3^*{}^2 a_2+a_1 a_2 a_3 -\tfrac{b_1 a_2}{2}-\tfrac{b_2 a_2}{2}-b_3 a_2+5 b_5 a_2-4 b_6 a_2+4 b_7 a_2-5 a_1 a_3^* a_2+b_5^* a_2\\\non
  &-\tfrac{1}{2} a_1 a_2^*{}^2-a_3 a_2^*{}^2-2 a_1 a_3^*{}^2-4 a_3 a_3^*{}^2-2 |a_2|^2 a_1-2 |a_3|^2 a_1-4 |a_2|^2 a_3-4 |a_3|^2 a_3\\\non
  &+a_1 b_3-2 a_3 b_3-a_1 b_4+2 a_3 b_5-4 a_3 b_6+2 a_3 b_7-2 |a_3|^2 a_1^*+\tfrac{b_1 a_1^*}{2}+b_7 a_1^*-\tfrac{3 |a_1|^2 a_2^*}{2}\\\non
  &-4 |a_3|^2 a_2^*-2 a_3^2 a_2^*-a_1 a_3 a_2^*-\tfrac{b_1 a_2^*}{2}+\tfrac{b_2 a_2^*}{2}+b_7 a_2^*+2 b_8 a_2^*-3 a_3 a_1^* a_2^*-|a_1|^2 a_3^*\\\non
  &-3 |a_2|^2 a_3^*-b_1 a_3^*+b_2 a_3^*+2 b_8 a_3^*-2 a_1 a_2^* a_3^*+\tfrac{a_1 b_1^*}{2}+a_3 b_1^*+\tfrac{a_1 b_2^*}{2}+a_3 b_2^*+a_1 b_7^*+2 a_3 b_7^* \ .
\end{align}\end{subequations}}%

\section{Lagrange multiplier analysis of the KS action}\label{sect:LagMul}

As argued in \cite{Casalbuoni1989,KS}, 
the KS model defined by the constraint \eqref{f^2}
and the action \eqref{eqn:KS0}
may be described in terms of two unconstrained chiral scalars, 
a dynamical superfield $\F$ and a Lagrange multiplier chiral superfield $\L$.
Their dynamics are governed by the action 
\begin{align} \label{eqn:KS0Lag}
	S[\F,\bar\F;\L,\bar\L] = \intx {\rm d}^4\q\, \bar\F\F 
	+ \left(\intx {\rm d}^2\q\, \Big(f\F+\frac12\L\F^2\Big) + \cc\right) \ ,
\end{align}
and the corresponding equations of motion are
\begin{subequations} \label{eqn:KS0LagEOM}
\begin{align}
 \label{eqn:KS0LagEOM1}
	\frac{\d S}{\d\L} &= \F^2 = 0 \,,  \\
 \label{eqn:KS0LagEOM2}
	\frac{\d S}{\d\F} &= -\frac14\Db^2\Fb + f + \L\F = 0 \ .
\end{align}
\end{subequations}
It appears that the relationship of the model (\ref{eqn:KS0Lag}) to the KS one
requires some more analysis than the brief discussions
given in \cite{Casalbuoni1989,KS}. The subtlety is that 
the Lagrange multiplier $\L$ can not be determined in terms of $\F$ and $\bar\F$ 
on the mass shell, due to the existence of an on-shell gauge invariance 
\begin{equation}
	\L \to \L + X \F\,, 
\end{equation}
for an arbitrary chiral scalar $X$. 
So it is not immediately obvious that the equations of motion yield a unique
Goldstino dynamics equivalent to that of the KS model.
As a result, it  is necessary to show that the part of $\L$ that is not fixed
by the equations of motion decouples from
the dynamics of the Goldstino dictated by \eqref{eqn:KS0LagEOM}.%
\footnote{As noted in \cite{KS}, 
multiplying the equation \eqref{eqn:KS0LagEOM2} by $\Phi$
and using the nilpotency condition \eqref{eqn:KS0LagEOM1}
yields Ro\v{c}ek's second constraint \eqref{eqn:Roceks_Constraint2}.
To prove the equivalence of \eqref{eqn:KS0Lag} to Ro\v{c}ek's model,
one still has to show that
(i) $\L$ decouples from the Goldstino;
(ii) the equation (\ref{eqn:KS0LagEOM2}) implies
the equation of motion for the Goldstino which follows from (\ref{eqn:ActionR}).
}

With the motivation given, 
let us perform the component analysis of the equations 
(\ref{eqn:KS0LagEOM1}) and  (\ref{eqn:KS0LagEOM2}). 
The component fields of $\F$ and $\L$ are introduced in the standard way:
\begin{align}
	\F(\q,\bar\q) &= \rme^{\rmi\q\s^a\bar\q\pd_a} 
						\Big(\f+\sqrt2\q\j+\q^2F \Big)\,,&
	\L(\q,\bar\q) &= \rme^{\rmi\q\s^a\bar\q\pd_a} 
						\Big( \m+\sqrt2\q\c+\q^2G \Big) \ .
\end{align}
The component equations of motion generated 
by \eqref{eqn:KS0LagEOM1} are all solved by
$\f = \frac12 F^{-1}\j^2$.
Then, the component equations of motion 
encoded in \eqref{eqn:KS0LagEOM2} become
\begin{subequations}\label{eqn:KSLagCompEOM}\begin{align} 
  0 &= \bar{F} + f + F^{-1}\m\j^2 	\,, 					\label{KSLagEOMa}\\
  0 &= \rmi (\pd\bar\j)_\a + 2\m\j_\a + F^{-1}\j^2\c_\a \,,	\label{KSLagEOMb}\\
  0 &= \frac12\Box(\bar F^{-1}\bar\j^2) + 2\m F - 2\c\j + F^{-1}\j^2G \ .		
															\label{KSLagEOMc}
\end{align}\end{subequations}
The equation of motion for $F$ \eqref{KSLagEOMa} 
can be iteratively solved for $F$
\begin{align} 
	F = -f (1 - f^{-2}\bar\m\bar\j^2 - f^{-4}\m\bar\m\j^2\bar\j^2) \ .
\end{align}
This solution can be substituted into \eqref{KSLagEOMc} which can then be
solved to yield 
\begin{align} 
	\m &= -f^{-1}\c\j \Big(1 - \frac14f^{-4} \bar\j^2\Box\j^2\Big)
		-\frac12f^{-2}G\j^2 \Big(1 - \frac12f^{-4} \bar\j^2\Box\j^2\Big)	
		+\frac1{16}f^{-6} \bar\j^2\Box\j^2\Box\bar\j^2  \non \\
		&-\frac14f^{-2}\Box\left(\bar\j^2
			\Big(1-\frac14f^{-4}\bar\j^2\Box\j^2\Big)\right)
		-\frac1{32}f^{-10}\j^2\bar\j^2\Box\j^2\Box\bar\j^2\Box\bar\j^2 \ ,
\end{align}
which is then substituted back into $F$ to give
\begin{align} 
	F = -f\left(1 + \frac14f^{-4}\bar\j^2\Box\j^2 
		- \frac1{16}f^{-8}(\j^2\bar\j^2\Box\j^2\Box\bar\j^2)\right) \ .
\end{align}
Note that this solution is free of $\c_\a$ and $G$.

The final equation of motion \eqref{KSLagEOMb} 
is the equation for the Goldstino dynamics. 
Substituting our solutions for $F$ and $\m$ into \eqref{KSLagEOMb} shows
that all $\c_\a$ and $G$ dependence cancels out and we a left with 
the dynamical equation  of the KS action \eqref{eqn:KS1}
\begin{align} 
	0 = \rmi(\pd\bar\j)_\a - \frac12 f^{-2} \j_a\Box\bar\j^2 
		+ \frac18 f^{-6} \left(\j_\a\bar\j^2\Box\j^2\Box\bar\j^2
		+ \j_\a\Box(\j^2\bar\j^2\Box\bar\j^2)\right)\ .
\end{align}
This means that the Lagrange multiplier fields $\c_\a$ and $G$ completely 
decouple from the Goldstino dynamics and may consistently be ignored.%
\footnote{In \cite{Casalbuoni1989}, these fields were set to zero 
and they worked with a gauge fixed action. 
Our consideration shows why the gauge conditions $\c_\a =0$ and $G=0$ 
are legitimate.
}
Due to the field redefinition \eqref{eqn:RtoKS} this is equivalent to
the equations of motions for Ro\v{c}ek's model \eqref{eqn:ActionR}.

In \cite{KS} Komargodski and Seiberg generalized their minimal Goldstino model, 
which is given by the constraint (\ref{f^2}) and the action \eqref{eqn:KS0},
to include higher-derivative interactions 
and couplings to supersymmetric matter.
In all of their models, 
the  Goldstino is always described by a chiral scalar $\F$ subject 
to the same constraint (\ref{f^2}). 
This constraint can be incorporated into the action 
at the cost of introducing a Lagrange multiplier chiral superfield $\L$, 
in complete analogy with the model (\ref{eqn:KS0Lag}).
It would be interesting to analyse the on-shell decoupling 
of the Lagrange multiplier for these more general theories.



\begin{footnotesize}
\providecommand{\href}[2]{#2}
\begingroup\raggedright
\endgroup
\end{footnotesize}

\end{document}